\documentclass[a4paper,11pt]{article}
\usepackage{jheppub}
\usepackage{caption}
\usepackage{setspace}
\usepackage{hyperref}
\usepackage{subfig}
\title{\boldmath Spin-dependent Regge-Wheeler Potential and QNMs in Quantum Corrected AdS Black Hole with Phantom Global Monopoles}

\author[a]{Faizuddin Ahmed} 
\author[b,1]{,\,\,Ahmad Al-Badawi} 
\author[c]{and\,\,\.{I}zzet Sakall{\i}}

\affiliation[a]{Department of Physics, University of Science \& Technology Meghalaya, Ri-Bhoi, Meghalaya, 793101, India}
\affiliation[b,1]{Department of Physics, Al-Hussein Bin Talal University, 71111, Ma’an, Jordan}
\affiliation[c]{AS245 Department of Physics, Eastern Mediterranean University, 99628, Famagusta Northern Cyprus, via Mersin 10, Turkiye}

\emailAdd{faizuddinahmed15@gmail.com}
\emailAdd{ahmadbadawi@ahu.edu.jo}
\emailAdd{izzet.sakalli@emu.edu.tr}

\abstract{
In this paper, we investigate the geodesic motion of test particles in the spacetime surrounding a static, spherically symmetric black hole, which is described by an AdS-Schwarzschild-like metric and incorporates a quantum correction. This black hole also features phantom global monopoles, which modify the structure of the black hole space-time. We begin by deriving the effective potential governing the motion of test particles in this system and carefully analyze the impact of quantum correction in the presence of both phantom and ordinary global monopoles. Furthermore, we extend our study to include the spin-dependent Regge-Wheeler (RW) potential, which characterizes the dynamics of perturbations in this quantum-corrected black hole background. By examining this RW potential for various spin fields, we show how quantum corrections affect its form in the presence of both phantom and ordinary global monopoles. Our analysis demonstrate that quantum correction significantly alter the nature of the RW-potential, influencing the stability and behavior of test particles and perturbations around the black hole.
}


\begin{document}
\maketitle
\flushbottom

\section{Introduction}

A long time ago, General Relativity (GR) theoretically predicted the existence of interesting and fascinating black holes (BHs). These enigmatic objects are solutions to Einstein's field equations and exhibit extreme curvature effects that profoundly influence their surrounding spacetime. In 2016, LIGO made the first direct detection of gravitational waves caused by the merger of binary BHs, providing solid observational evidence for GR \cite{ligo}. This groundbreaking discovery has sparked an unprecedented level of interest in BH physics, both from theoretical and observational perspectives. Recently, the existence of BHs was proved by the Event Horizon Telescope (EHT) collaboration when it first captured the shadow of $M87^{*}$, a supermassive BH in the nearby galaxy \cite{AA3}. These observations remain insufficient, as several fundamental properties of BHs, including Hawking radiation \cite{Hawking:1974rv,Gibbons:1977mu,Hawking:1982dh,Parikh:1999mf,Sakalli:2017ewb,Sakalli:2015jaa}, entropy \cite{Ryu:2006bv,Strominger:1996sh,Bardeen:1973gs}, and other thermodynamic characteristics \cite{Eisert:2008ur,Bousso:2002ju}, still require further investigation and empirical validation. However, despite the remarkable success of GR, the Schwarzschild BH solution exhibits a significant flaw: a singularity at $r = 0$, where the curvature of spacetime becomes infinite. This intrinsic singularity hinders a complete description of the background structure, raising profound issues about causality and predictability \cite{ph1,ph2}.
 
Numerous approaches have been proposed in the literature to address this singularity problem \cite{ph3,DIK,ph5,ph6,ph7,ph8,ph9,ph10}. One promising direction involves incorporating quantum phenomena, as quantum mechanics provides a natural framework to regulate physical quantities near such singularities. Specifically, quantum corrections can act as a barrier to the formation of singularities at the origin \cite{ph12,ph13}. The first quantum-corrected BH solution was introduced in Ref. \cite{ph11}. This pioneering work has been followed by extensive investigations into the role of quantum effects in BH spacetimes, including studies on event horizons, thermodynamics, and spherically symmetric gravitational collapses. A particularly influential model was proposed by Kazakov and Solodukhin \cite{DIK}, who introduced a regular spacetime free of singularities by incorporating spherically symmetric quantum fluctuations of the metric and matter fields. This approach led to an effective two-dimensional dilaton gravity model, providing a valuable platform for further theoretical developments.

Quasinormal modes (QNMs) are a fundamental aspect of BH perturbation theory, describing the oscillatory response of a BH's spacetime to external perturbations \cite{CC1,CC2,Chen:2021cts,MB2,Moderski:2001tk}. Unlike normal modes, QNMs do not persist indefinitely due to energy dissipation, which occurs through flux conservation, causing the oscillations to decay over time. These modes are typically modeled using an exponential form, $e^{-i\,\omega\,t}$, where $\omega$ is a complex frequency. The real part of the frequency, $\mathrm{Re}(\omega)$, corresponds to the oscillation frequency, while the imaginary part, $\mathrm{Im}(\omega)$, represents the decay rate of the oscillations. This complex frequency structure is what gives the modes their "quasi-normal" character, as the oscillations eventually fade due to the loss of energy. Mathematically, QNMs arise from the linearized equations of GR, which describe perturbations around a BH solution. The perturbations lead to a second-order linear partial differential equation known as the master equation. These equations describe the evolution of various types of perturbations-gravitational, electromagnetic, or scalar-around BH backgrounds, such as Schwarzschild, Kerr, or Reissner-Nordstr\"{o}m BHs. Finding exact solutions to these equations is generally difficult due to their complexity. Consequently, a variety of approximate methods have been developed to study QNMs in different BH configurations, including the WKB approximation, the asymptotic iteration method, continued fraction techniques, and direct numerical methods for solving the equations of motion (see, for example, Refs. \cite{CC3, CC4, CC5, CC6, CC7, CC8, CC9, CC10, CC11, CC12, CC13, CC14, CC15}).

The significance of QNMs has grown substantially in recent years due to their connection with astrophysical observations \cite{Zahid:2024hwi,Balart:2024rtj,Al-Badawi:2024jnt,Al-Badawi:2024cby,Al-Badawi:2024iqv,Al-Badawi:2024iax}. The detection of gravitational waves by the LIGO and Virgo collaborations \cite{Sasaki:2016jop} has opened a new window into BH physics. Gravitational wave signals from events like binary BH mergers contain valuable information about the properties of the BHs involved, including their mass, spin, and QNM frequencies. By analyzing these frequencies, researchers can infer key characteristics of the BHs and test GR in extreme gravitational fields. The comparison of theoretical QNM spectra with observed gravitational wave data has proven to be an effective tool for extracting information about the nature of BHs, further solidifying the importance of QNMs in modern astrophysical research.

In light of the above considerations, we are interested in studying the geodesic and RW potential as well as QNMs of a static, spherically symmetric BH described by an AdS-Schwarzschild-like metric with quantum correction. The line-element describing a spherically symmetric BH with global monopoles (ordinary and phantom) in the spherical coordinates $(t, r, \theta, \phi)$ is given by \cite{SCJJ, AHEP}
\begin{equation}
ds^2=-\mathcal{A}(r)\,dt^2+\frac{dr^2}{\mathcal{A}(r)}+r^2\,(d\theta^2+\sin^2 \theta\,d\phi^2), \label{a1}
\end{equation}
with the function
\begin{equation}
    \mathcal{A}(r)=\Big(1-8\,\pi\,\eta^2\,\xi-\frac{2\,M}{r}\Big),\label{a2}
\end{equation}
where $\eta, M$ and $\xi$ stand for the energy scale of symmetry breaking, the quantity of matter, and BH kinetic energy, respectively. The scenario mediated by $\xi=1$ describes an arena of an ordinary global monopole that arises from the scalar field’s non-negative and non-zero kinetic energy \cite{MB,ERBM}. However, the phantom global monopole is generated by selecting $-1$ value of $\xi$, thereby relating it with the scalar field’s negative kinetic energy.\\
On examining the quantum fluctuations associated with a spherically symmetric manifold, significant corrections in the framework of 2D dilaton gravity theory emerge, mediated by the Einstein-Hilbert action combined with 4D interaction theory \cite{DIK}. Such a theoretical setting enables the renormalization of gravitational theories like the 2D dilaton gravity model initially proposed by Kazakov and Solodukhin. In \cite{wu}, the authors extended this approach by presenting a quantum-corrected AdS-Schwarzschild black hole (BH) and analyzing its thermodynamic properties in detail. A pivotal aspect of black hole physics is the examination of perturbations, which are typically described by the Regge-Wheeler (RW) potential. This potential encapsulates the dynamics of scalar, electromagnetic, and gravitational perturbations within BH spacetimes, providing a robust framework for studying quasinormal modes (QNMs). In the quantum-corrected metric addressed in this analysis, notable alterations to the RW potential are introduced. These modifications are intricately influenced by parameters such as the monopole's nature (ordinary or phantom) and the quantum correction parameter $\alpha$, thereby offering deeper insights into the interplay between quantum effects and classical BH perturbations. The primary objective of this study is to analyze the RW potential derived in the background of an AdS BH \cite{wu}, incorporating quantum corrections in the presence of both phantom and ordinary global monopoles. As mentioned previously, the RW potential governs the dynamics of perturbations for scalar fields (spin-0), electromagnetic fields (spin-1), and gravitational fields (spin-2) \cite{Birrell:1982ix,Sakalli:2022xrb}. In this work, we examine how quantum corrections affect the RW potential for these different spin fields. Additionally, we investigate how the presence of phantom and ordinary global monopoles modifies the RW potential, comparing these results with those obtained for the Schwarzschild BH. 

This paper is organized as follows: In Sec. \ref{sec2}, we introduce the quantum-corrected AdS-Schwarzschild BH with global monopoles and examine its fundamental properties, including its horizon structure and modifications to spacetime geometry. Section \ref{sec3} focuses on the geodesic motion of test particles and photons, providing a comprehensive analysis of the effective potential and its dependence on key parameters. In Sec. \ref{sec4}, we analyze the RW potential for different perturbation types and calculate quasinormal frequencies, emphasizing the influence of quantum corrections and monopole effects. Finally, in Sec. \ref{sec5}, we summarize our findings and discuss potential avenues for future research.

\section{Quantum-Corrected AdS BH with Phantom Global Monopoles} \label{sec2}

Inspired by the above discussion, now we are going to introduce a line-element ansatz of a spherically symmetric BHs with ordinary and phantom global monopole taking into account the quantum correction given by
\begin{eqnarray}
ds^2=-\mathcal{F}(r)\,dt^2+\frac{dr^2}{\mathcal{F}(r)}+r^2\,(d\theta^2+\sin^2 \theta\,d\phi^2).\label{a5}
\end{eqnarray}
where $\mathcal{F}$ is given by
\begin{equation}
    \mathcal{F} (r)=\frac{1}{r}\,\sqrt{r^2-\alpha^2}-\frac{2\,M}{r}-8\,\pi\,\eta^2\,\xi-\frac{\Lambda}{3}\,\frac{(r^2-\alpha^2)^{3/2}}{r}, \label{a6}
\end{equation}
in which  $\Lambda=-\frac{3}{\ell^2}$ is the cosmological constant and $\alpha=4\,\ell_{p}$ is a quantum (small) correction that describes the behavior of spherical symmetric quantum fluctuations. If we disregard the AdS background (i.e., take $\Lambda \to 0$), the space-time described by equation (\ref{a5}) reduces to a deformation or quantum fluctuation of the Schwarzschild metric, as obtained in Ref. \cite{DIK}. 
 From the above expression, it is evident that $r \geq \alpha=4\,\ell_p$, which suggests the existence of a minimum length scale in spacetime. This is a characteristic feature often associated with quantum gravity theories. Since $\alpha$ is a small correction term, the series expansion of $\alpha$ in the function (\ref{a6}) results
\begin{equation}
    \mathcal{F} (r) \simeq 1-8\,\pi\,\eta^2\,\xi-\frac{2\,M}{r}-\frac{\Lambda}{3}\,r^2+\left(\frac{\Lambda}{2} -\frac{1}{2\,r^2}\right)\,\alpha^2+\mathcal{O} (\alpha^4).\label{a7}
\end{equation}
Throughout this manuscript, we will use the space-time (\ref{a5}) with the function (\ref{a7}) and discuss our objectives in detail.

It is clear that, in comparison with the usual AdS-Schwarzschild BH metric with phantom global monopoles, quantum corrections introduce additional terms, specifically $(-\alpha^2/2r^2)$ and $(\Lambda \alpha^2 / 2)$. The former term suggests that quantum fluctuations contribute to the effect of a virtual charge. If we ignore the AdS background (i.e., set $\Lambda \to 0$), the line element (\ref{a5}) with the function (\ref{a7}) for $\xi = 1$ becomes analogous to the global monopole spacetime presented in Ref. \cite{MB,ERBM}, where a quantum potential similar to the one discussed in Ref. \cite{AFA} is considered.

The event horizon of the BH solution (\ref{a5}) exist at a radius given by (setting zero cosmological constant)
\begin{eqnarray}
    r_h=\frac{M\,\left(1+\sqrt{1+\frac{\alpha^2}{2\,M^2}\,(1-8\,\pi\,\eta^2\,\xi)}\right)}{(1-8\,\pi\,\eta^2\,\xi)},\label{a9}
\end{eqnarray}
For $\xi=1$ which corresponds to ordinary global monopole, the horizon radius will be
\begin{eqnarray}
    &&r_h=\frac{M\,\left(1+\sqrt{1+\frac{\alpha^2\,\mathrm{a}}{2\,M^2}}\right)}{\mathrm{a}}\simeq \frac{2\,M}{\mathrm{a}}+\frac{\alpha^2}{4\,M},\quad\quad \mathrm{a}=(1-8\,\pi\,\eta^2),\label{a10}\\
    &&r_h=M\,\left(1+\sqrt{1+\frac{\alpha^2}{2\,M^2}}\right) \simeq 2\,M+\frac{\alpha^2}{4\,M}\quad\quad:\mbox{deformed-Schwarzschild metric},\label{a11}\\
    &&r_h=2\,M\quad\quad:\mbox{Schwarzschild metric}.\label{a12}
\end{eqnarray}
We see that the horizon radius will be $r_{h} > r^{\mbox{deformed-Sch}}_{h} > r^{\mbox{Sch}}_{h}$. 

For $\xi=-1$ which corresponds to phantom global monopole, the horizon radius will be
\begin{eqnarray}
    r_h=\frac{M\,\left(1+\sqrt{1+\frac{\alpha^2\,\mathrm{a}}{2\,M^2}}\right)}{\mathrm{b}} \simeq \frac{2\,M}{\mathrm{b}}+\frac{\alpha^2}{4\,M},\quad \mathrm{b}=(1+8\,\pi\,\eta^2).\label{a13}
\end{eqnarray}
Thus, the presence of phantom global monopole will change the horizon radius as $r_{h} < r^{\mbox{Sch}}_{h} < r^{\mbox{deformed-Sch}}_{ph}$. Overall we can say that the presence of global monopole in deformations of the Schwarzschild-metric increases the horizon radius compared to phantom global monopole.

\section{Particle Dynamics} \label{sec3}

In this section, we study geodesics motions of test particles around this BH solution (\ref{a5}) and analyze the outcomes. The Lagrangian function using the space-time (\ref{a5}) at the equatorial plane defined by $\theta=\pi/2$ is given by
\begin{equation}
    \mathcal{L}=\frac{1}{2}\,\left[-\mathcal{F}(r)\,\dot{t}^2+\frac{\dot{r}^2}{\mathcal{F}(r)}+r^2\,\dot{\phi}^2\right].\label{bb1}
\end{equation}

We see from the space-time (\ref{a5}) that the metric tensor $g_{\mu\nu}$ is independent of the coordinates $(t, \phi)$, while depends on $(r, \theta)$ coordinates. Hence, there are two Killing vectors, namely $\partial_{t}$ and $\partial_{\phi}$. The constant of motion associated with these Killing vectors are defined by
\begin{eqnarray}
    &&-\mathrm{E}=-\mathcal{F}(r)\,\dot{t}\Rightarrow \dot{t}=\frac{\mathrm{E}}{\mathcal{F}(r)},\label{bb2}\\
    &&\mathrm{L}=r^2\,\dot{\phi}\Rightarrow \dot{\phi}=\frac{\mathrm{L}}{r^2},\label{bb3}
\end{eqnarray}
where $\mathrm{E}$ is the particles energy and $\mathrm{L}$ is the conserved angular momentum.

Substituting $\dot{t}$ and $\dot{\phi}$ into the equation (\ref{bb1}) results the following equation
\begin{equation}
    \dot{r}^2+V_{eff} (r)=\mathrm{E}^2\label{bb4}
\end{equation}
which represents one-dimensional equation of motions of a test particles having energy $\mathrm{E}$ and the effective potential $V_{eff}$ given by
\begin{equation}
    V_{eff} (r)=\left(-\epsilon+\frac{\mathrm{L}^2}{r^2}\right)\,\left[1-\frac{2\,M}{r}-8\,\pi\,\eta^2\,\xi-\frac{\Lambda}{3}\,r^2+\left(\frac{\Lambda}{2}-\frac{1}{2\,r^2}\right)\,\alpha^2\right].\label{bb9}
\end{equation}

Setting $\xi=1$ which corresponds to ordinary global monopole, the effective potential of the system will be
\begin{equation}
    V_{eff} (r)=\left(-\epsilon+\frac{\mathrm{L}^2}{r^2}\right)\,\left[\mathrm{a}-\frac{2\,M}{r}-\frac{\Lambda}{3}\,r^2+\left(\frac{\Lambda}{2}-\frac{1}{2\,r^2}\right)\,\alpha^2\right].\label{bb9a}
\end{equation}

For $\xi=-1$ which corresponds to phantom global monopole, the effective potential of the system will be
\begin{equation}
    V_{eff} (r)=\left(-\epsilon+\frac{\mathrm{L}^2}{r^2}\right)\,\left[\mathrm{b}-\frac{2\,M}{r}-\frac{\Lambda}{3}\,r^2+\left(\frac{\Lambda}{2}-\frac{1}{2\,r^2}\right)\,\alpha^2\right].\label{bb9b}
\end{equation}
From expressions (\ref{bb9a}) and (\ref{bb9b}), it is evident that the effective potential for the system with an ordinary global monopole is lower than that for the system with a phantom global monopole.

We have generated Figures \ref{fig:13} to \ref{fig:14} illustrating the effective potential of the system, as described by Eq. (\ref{bb9}), for both null ($\epsilon = 0$) and time-like ($\epsilon = -1$) geodesics. These plots show the effective potential as a function of the energy scale $\eta$ for various values of the parameters ($\alpha, \Lambda$). Dotted lines represent the case with an ordinary global monopole, while solid colored lines correspond to the case with a phantom global monopole.
\begin{center}
\begin{figure}[ht!]
\subfloat[$\alpha=0.1,\mathrm{L}=1$]{\centering{}\includegraphics[scale=0.32]{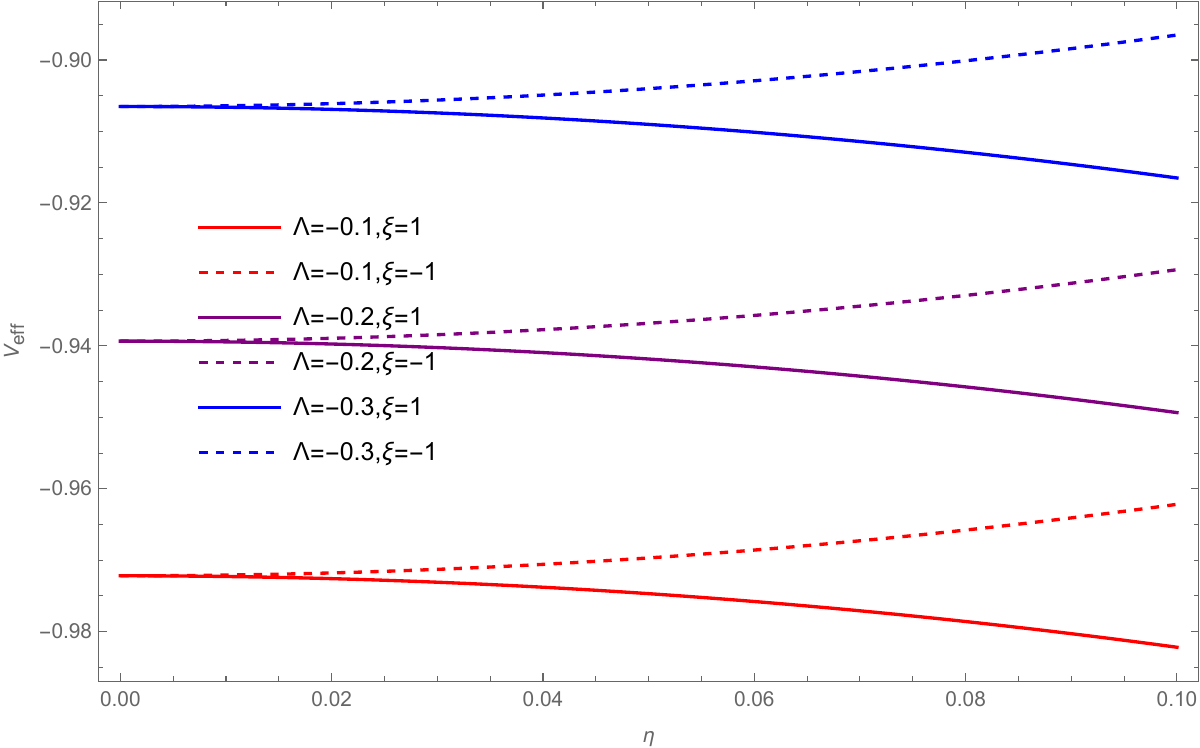}}\quad\quad
\subfloat[$\mathrm{L}=1$]{\centering{}\includegraphics[scale=0.32]{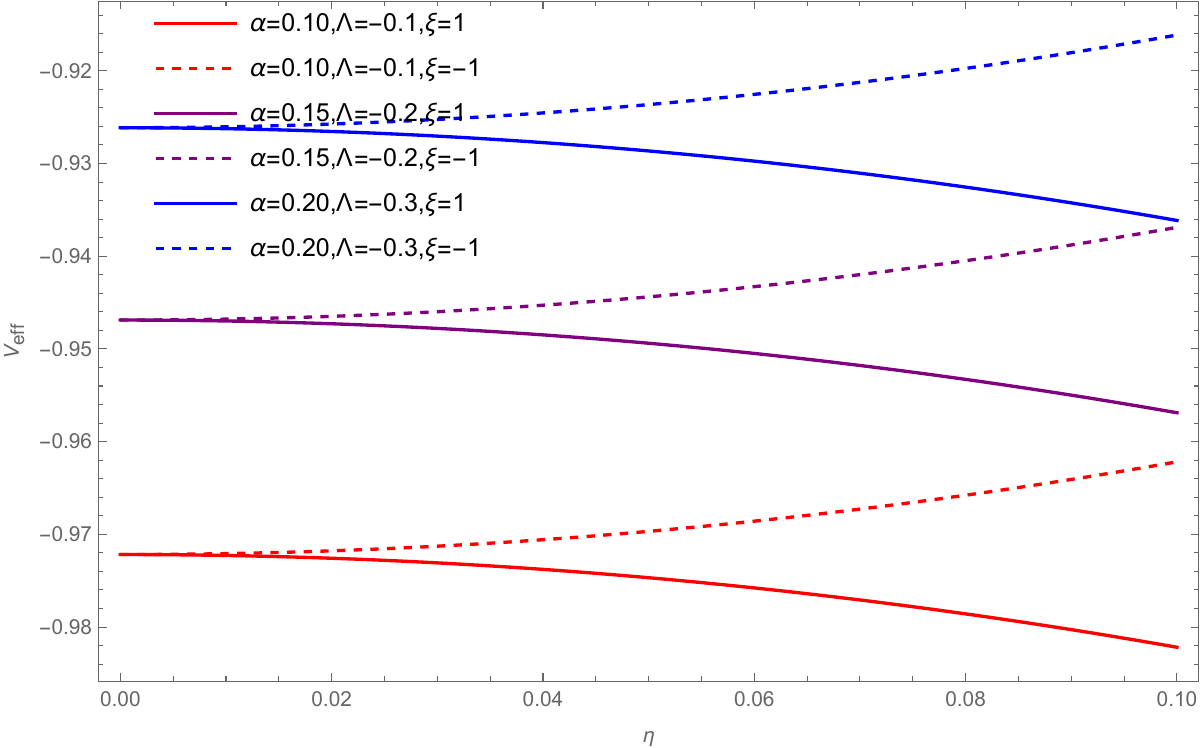}}
\hfill\\
\begin{centering}
\subfloat[$\Lambda=-0.3,\mathrm{L}=1$]{\centering{}\includegraphics[scale=0.32]{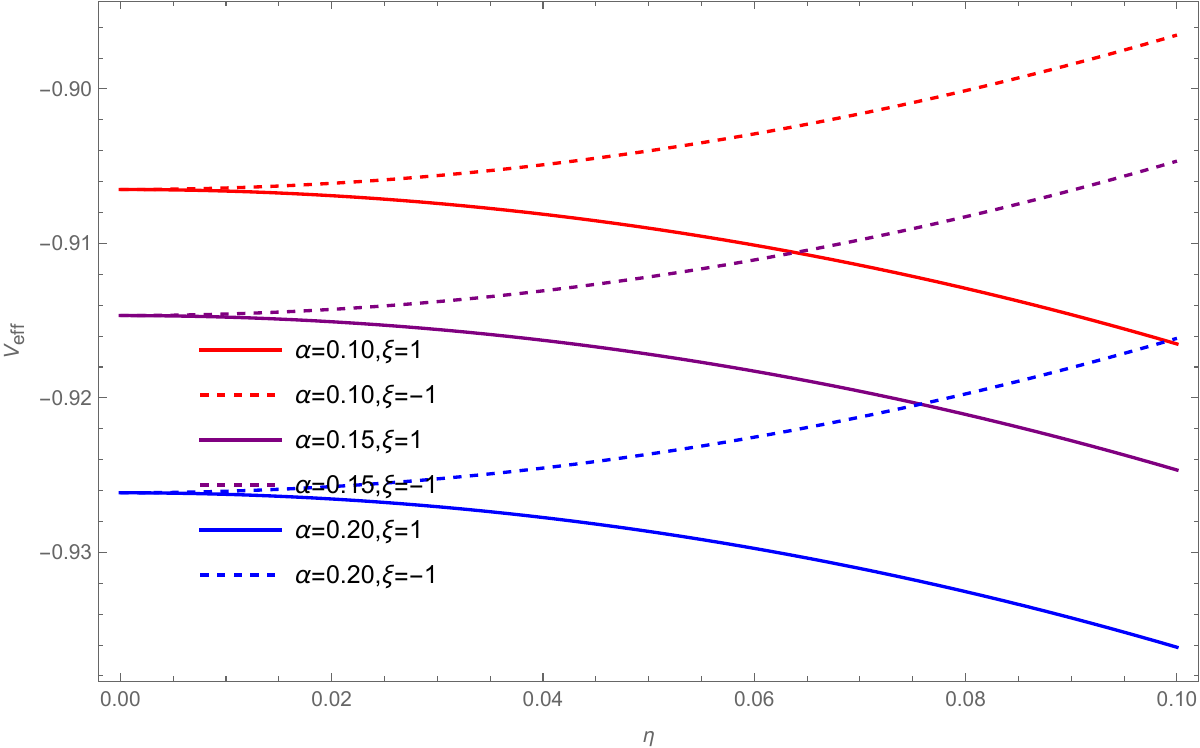}}
\end{centering}
\centering{}\caption{Effective potential with an ordinary and a phantom global monopole for null geodesics.}\label{fig:13}
\hfill\\
\subfloat[$\alpha=0.1,\mathrm{L}=1$]{\centering{}\includegraphics[scale=0.32]{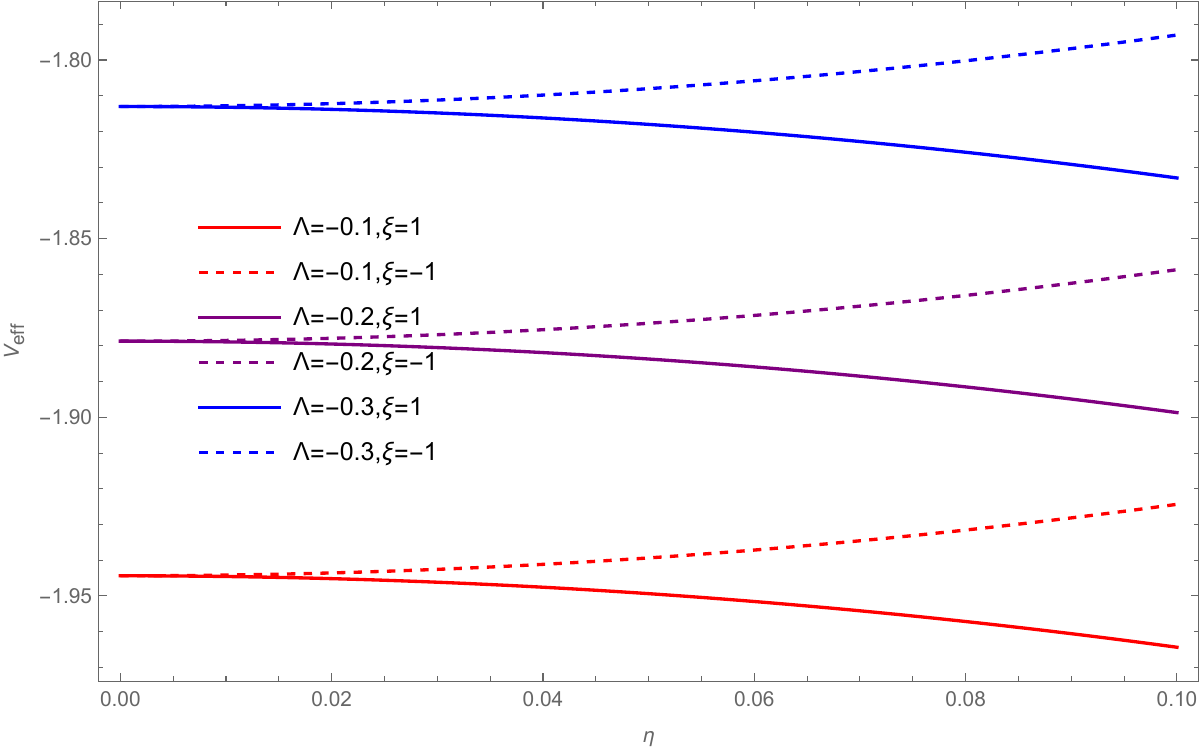}}\quad\quad
\subfloat[$\mathrm{L}=1$]{\centering{}\includegraphics[scale=0.32]{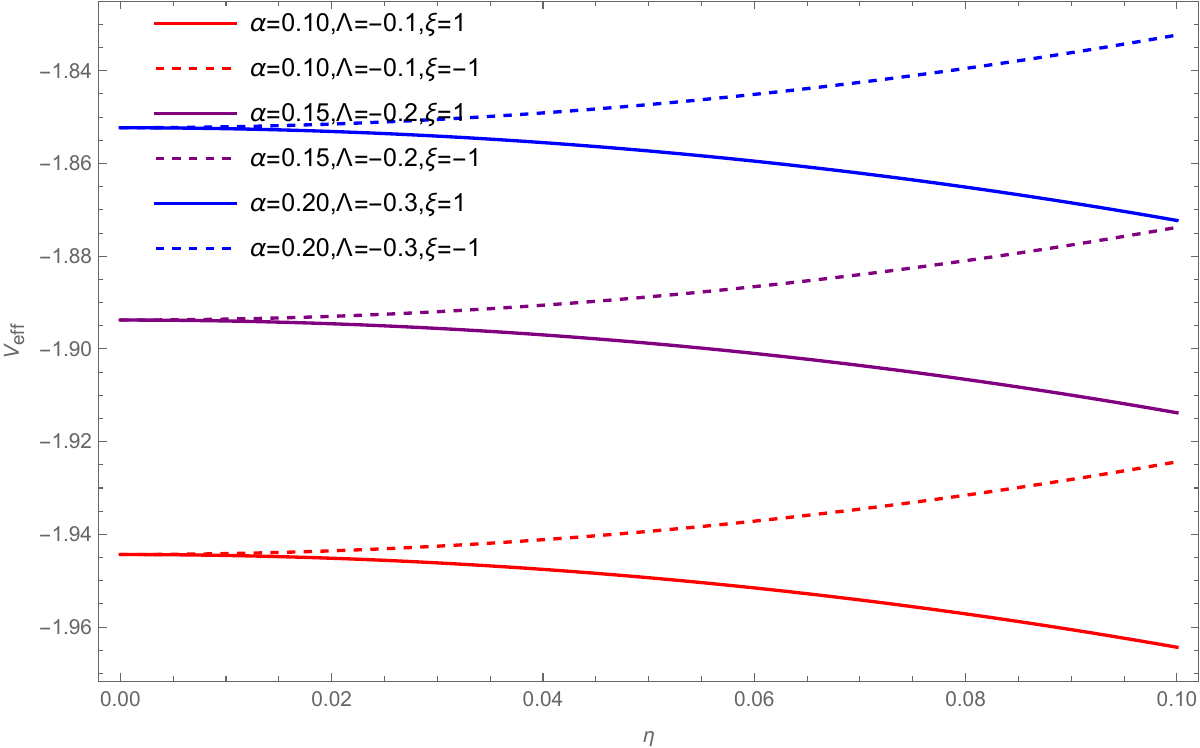}}
\hfill\\
\begin{centering}
\subfloat[$\Lambda=-0.3,\mathrm{L}=1$]{\centering{}\includegraphics[scale=0.32]{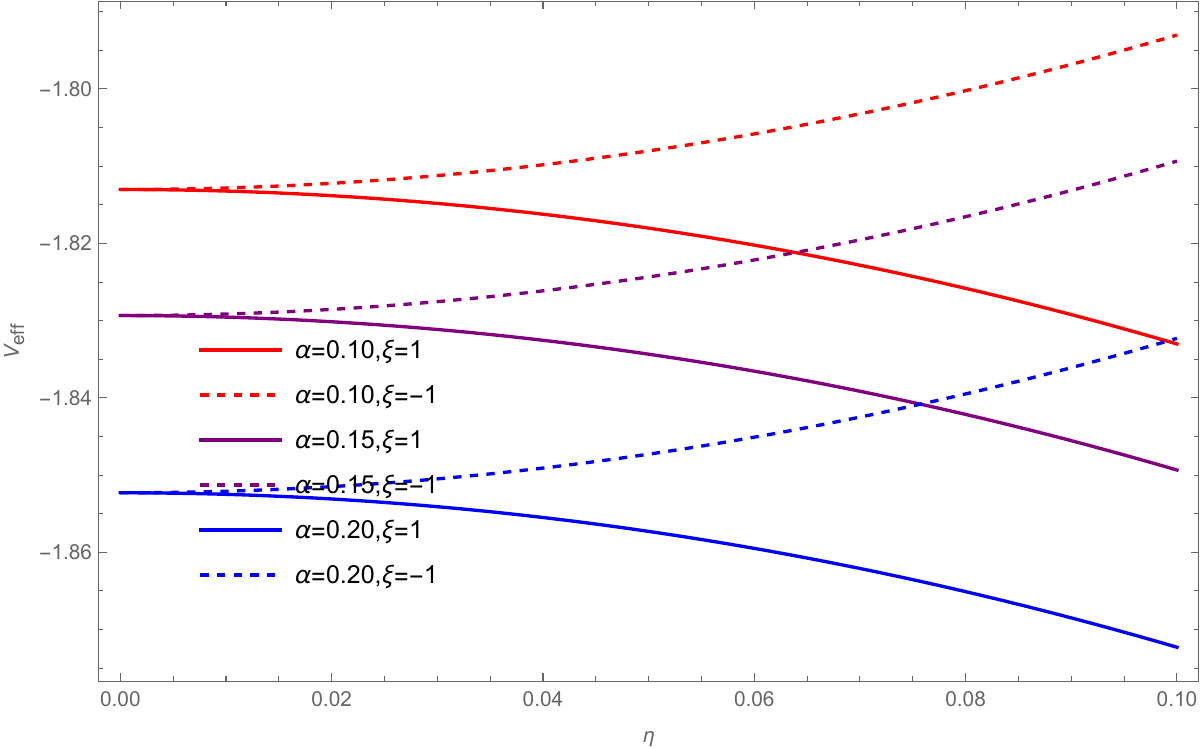}}
\end{centering}
\centering{}\caption{Effective potential with an ordinary and a phantom global monopole for time-like geodesics.}\label{fig:14}
\end{figure}
\par\end{center}
To determine the radius of photon sphere, we have
\begin{equation}
    \frac{dV_{eff}}{r}\Big{|}_{r=r_{ph}}=0.\label{bb10}
\end{equation}
Using expression (\ref{bb9}) for mull geodesics ($\epsilon=0$), we find the radius of photon sphere as follow (setting zero cosmological constant):
\begin{equation}
    r_{ph}=\frac{3\,M\,\left(1+\sqrt{1+\frac{2\,\alpha^2}{3\,M^2}\,(1-8\,\pi\,\eta^2\,\xi)}\right)}{2\,(1-8\,\pi\,\eta^2\,\xi)}\label{bb11}
\end{equation}
which reduces to the known result for the deformed-Schwarzschild metric obtained in \cite{DIK} provided $\xi \to 0$ here. This result in the limit $\alpha \to 0$ further reduced for the Schwarzschild metric. Thus, we can say that the presence of phantom global monopoles change the size of the photon sphere compared to the known result.

For $\xi=1$ which corresponds to ordinary global monopole, the photon sphere radius becomes
\begin{eqnarray}
    &&r_{ph}=\frac{3\,M\,\left(1+\sqrt{1+\frac{2\,\alpha^2\,\mathrm{a}}{3\,M^2}}\right)}{2\,\mathrm{a}}\simeq \frac{3\,M}{\mathrm{a}}+\frac{\alpha^2}{2\,M},\quad \mathrm{a}=(1-8\,\pi\,\eta^2)\quad :\mbox{current metric},\label{bb12}\\
    &&r_{ph}=\frac{3\,M\,\left(1+\sqrt{1+\frac{2\,\alpha^2}{3\,M^2}}\right)}{2} \simeq 3\,M+\frac{\alpha^2}{2\,M}\quad:\mbox{deformed Schwarzschild metric \cite{DIK}},\label{bb13}\\
    &&r_{ph}=3\,M\quad\quad :\mbox{Schwarzschild metric},\label{bb14}
\end{eqnarray}
One can easily show that the radius of photon sphere will be $r_{ph} > r^{\mbox{d-Sch}}_{ph} > r^{\mbox{Sch}}_{ph}$.

For $\xi=-1$ which corresponds to phantom global monopole, we find
\begin{equation}
    r_{ph}=\frac{3\,M\,\left(1+\sqrt{1+\frac{2\,\alpha^2\,\mathrm{b}}{3\,M^2}}\right)}{2\,\mathrm{b}} \simeq \frac{3\,M}{\mathrm{b}}+\frac{\alpha^2}{2\,M},\quad \mathrm{b}=(1+8\,\pi\,\eta^2).\label{bb15}
\end{equation}
In that case the radius of photon sphere will be $r_{ph} < r^{\mbox{Sch}}_{ph} < r^{\mbox{d-Sch}}_{ph}$.

\subsection{Force on the massless photon particle}

In this part, we calculate the force acting on the massless photon particles under the gravitational effects. This force can be calculated using the relation
\begin{equation}
    \mathrm{F}_{ph}=-\frac{1}{2}\,\frac{dV_{eff}}{dr}.\label{cc1}
\end{equation}

Using the effective potential $V_{eff}$ given by the expression (\ref{bb9}), we find this force (setting zero cosmological constant)
\begin{eqnarray}
    \mathrm{F}_{ph}&=&\frac{\mathrm{L}^2}{r^2}\,\left(\frac{1}{r}\,(1-8\,\pi\,\eta^2\,\xi)-\frac{3\,M}{r^2}-\frac{\alpha^2}{r^3}\right).\label{cc3}
\end{eqnarray}

In the presence of an ordinary global monopole, the expression of force on the massless photon particle will be
\begin{eqnarray}
    \mathrm{F}_{ph}(\mbox{ordinary})=\frac{\mathrm{L}^2}{r^5}\,\left(\mathrm{a}\,r^2-3\,M\,r-\alpha^2\right).\label{cc4}
\end{eqnarray}
The presence of the global monopole parameter, denoted by $\mathrm{a}$, and the quantum correction, $\alpha$, both have an impact on the force experienced by a massless photon particle, in contrast to the result obtained in the Schwarzschild metric.

While in the presence of a phantom global monople, it will be
\begin{eqnarray}
    \mathrm{F}_{ph}(\mbox{phantom})=\frac{\mathrm{L}^2}{r^5}\,\left(\mathrm{b}\,r^2-3\,M\,r-\alpha^2\right).\label{cc5}
\end{eqnarray}

\begin{figure}[ht!]
    \centering
    \includegraphics[width=0.45\linewidth]{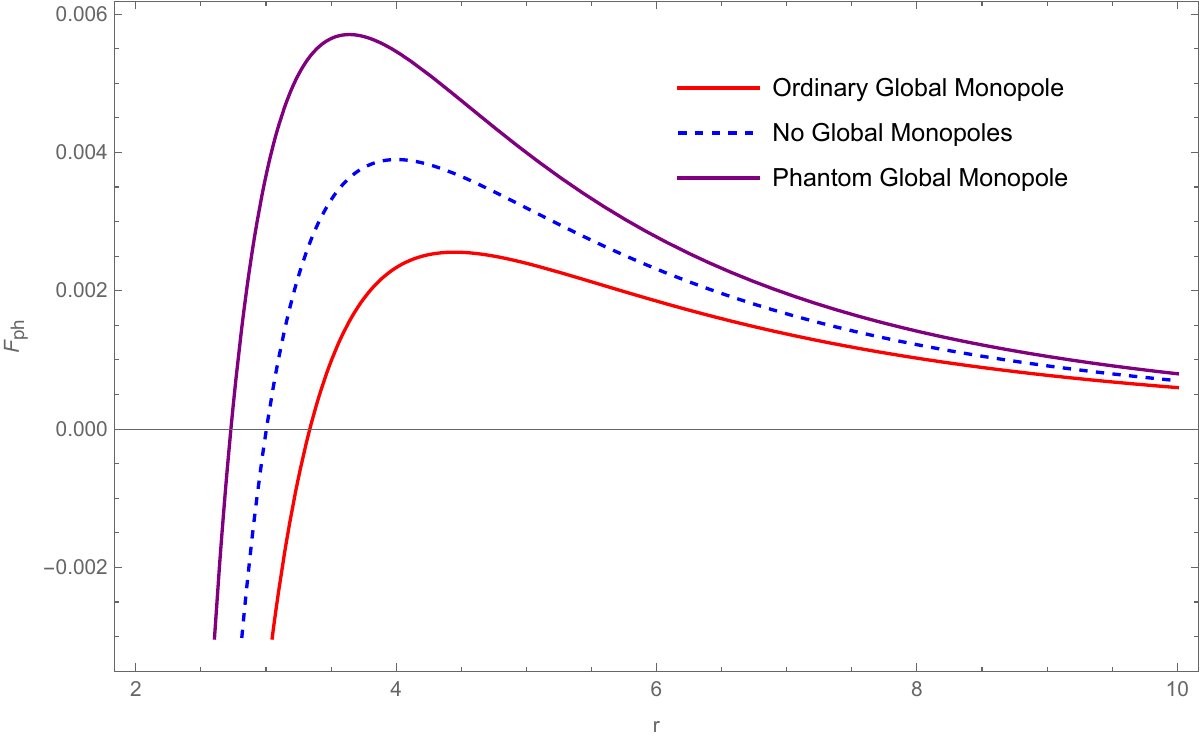}\quad\quad
    \includegraphics[width=0.45\linewidth]{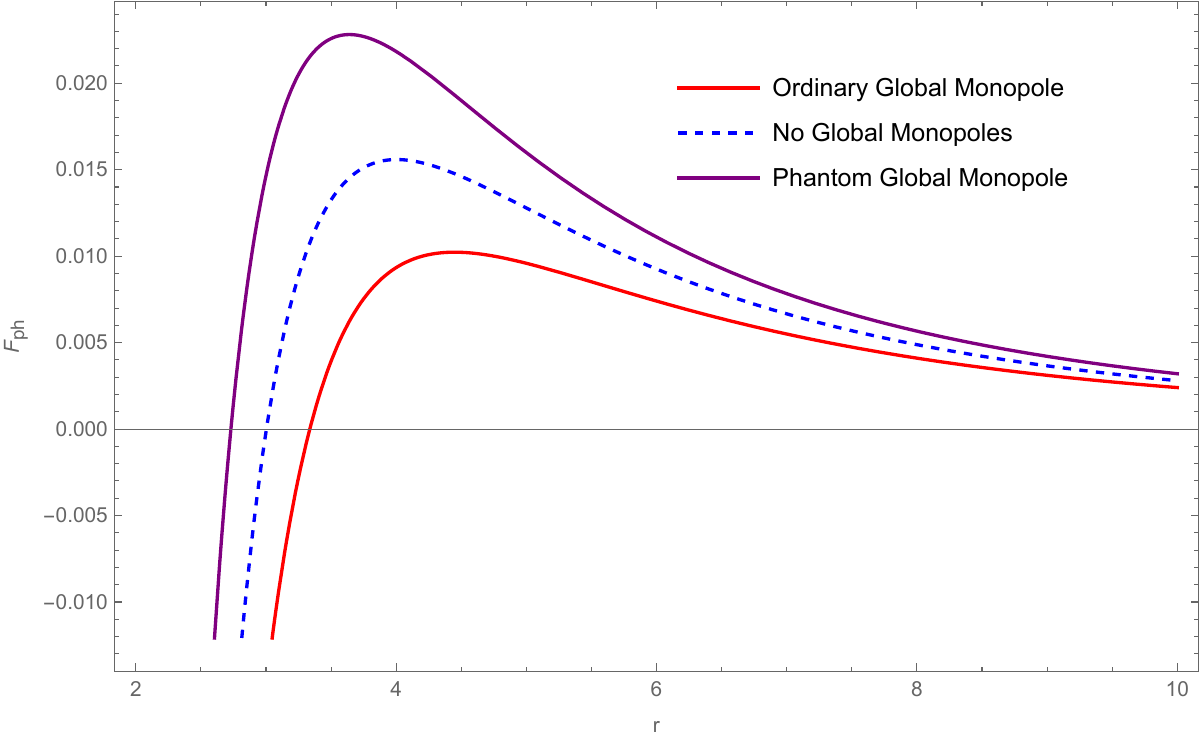}
    \caption{The force on the massless photon particle with and without global monopoles. Here, we set $M=1$, $\alpha=0.1$, $\mathrm{a}=0.9$, and $\mathrm{b}=1.1$. In the left panel, $\mathrm{L}=1$ and in the right panel, $\mathrm{L}=2$.}
    \label{fig:15}
\end{figure}

The presence of the phantom global monopole parameter, denoted by $\mathrm{b}$, along with the quantum correction, $\alpha$, both influence the force experienced by a massless photon particle, in contrast to the force observed in the Schwarzschild metric. Furthermore, it is evident that the $\mathrm{F}_{ph}(\mbox{phantom})$ associated with the phantom global monopole is greater than $\mathrm{F}_{ph}(\mbox{ordinary})$, which corresponds to the force in the presence of the ordinary global monopole since $\mathrm{b} > \mathrm{a}$. In Figure \ref{fig:15}, we show this force as a function of $r$ for various values of the conserved angular momentum $\mathrm{L}$, both in the presence and absence of global monopoles.

\section{The Regge-Wheeler potential} \label{sec4}

In this section, we investigate the spin-dependent RW potentials for the regular BH constructed in this paper. The RW-Zerilli equations describe the gravitational perturbations of a Schwarzschild-like BH in GR. These perturbations are categorized into two types: axial and polar perturbations. The equation governing axial perturbations is known as the RW equation \cite{TR}, while the equation for polar perturbations is referred to as the Zerilli equation \cite{FJZ}. Axial perturbations are typically more complex and do not lend themselves easily to the WKB approximation, making the computation of QNMs challenging without the use of numerical methods. Given this difficulty, we only examine the relevant RW potential for fields, such as zero-spin, spin-one and spin-two and analyze the result. 

To proceed for the RW-potential, we perform the following coordinate change (called tortoise coordinate) 
\begin{equation}
    dr_*=\frac{dr}{\mathcal{F}(r)},\quad\quad \partial_{r_*}=\mathcal{F}(r)\,\partial_r.\label{q2}
\end{equation}
into the line-element Eq. (\ref{a5}) results
\begin{equation}
    ds^2=\mathcal{F}(r_*)\,\{-dt^2+dr^2_{*}\}+\mathcal{H}^2(r_*)\,(d\theta^2+\sin^2 \theta\,d\phi^2),\label{qq3}
\end{equation}

In Regge and Wheeler’s original work \cite{TR}, they show that for perturbations in a BH space-time, assuming a separable wave form of the type
\begin{equation}
    \Phi(t, r_{*},\theta, \phi)=\exp(i\,\omega\,t)\,Y^{\ell}_{m} (\theta,\phi)\,\psi(r_*)/r_{*},\label{qq4}
\end{equation}
where $Y^{\ell}_{m} (\theta,\phi)$ are the spherical harmonics, $\omega$ is (possibly complex) temporal frequency in the Fourier domain \cite{TR}, and $\psi (r)$ is a propagating scalar, vector, or spin two axial bi-vector field in the candidate space-time. The RW equation is given by
\begin{equation}
    \frac{\partial^2 \psi(r_*)}{\partial r^2_{*}}+\left\{\omega^2-\mathcal{V}_s\right\}\,\psi(r_*)=0.\label{qq5}
\end{equation}
The method for solving Equation (\ref{qq5}) is dependent on the spin of the perturbations and on the background space-time.

The spin-dependent RW potential is given by the following expression
\begin{equation}
    \mathcal{V}_S=\frac{\mathcal{F}}{\mathcal{H}^2}\,\Big\{\ell\,(\ell+1)+S\,(S-1)\,(g^{rr}-1)\Big\}+(1-S)\,\frac{\partial^2_{r_{*}}\,\mathcal{H}}{\mathcal{H}},\label{qq6}
\end{equation}
where $\mathcal{F}$ and $\mathcal{H}$ are the relevant functions as specified by Equation (\ref{a5}), $\ell$ is the multipole number $\ell\geq s$, and $g^{rr}$ is the relevant contrametric component with respect to standard curvature coordinates (for which the covariant components are presented in Equation (\ref{a5})).

Therefore, the spin-dependent RW potential under consideration is given by
\begin{eqnarray}
    &&\mathcal{V}_S \simeq \left[1-8\,\pi\,\eta^2\,\xi-\frac{2\,M}{r}-\frac{\Lambda}{3}\,r^2+\left(\frac{\Lambda}{2} -\frac{1}{2\,r^2}\right)\,\alpha^2\right]\,\Bigg[\frac{\ell\,(\ell+1)}{r^2}+\frac{S\,(S-1)}{r^2}\times\nonumber\\
    &&\left(-8\,\pi\,\eta^2\,\xi-\frac{2\,M}{r}-\frac{\Lambda}{3}\,r^2+\left(\frac{\Lambda}{2} -\frac{1}{2\,r^2}\right)\,\alpha^2\right)
    +\frac{(1-S)}{r}\,\left(\frac{2\,M}{r^2}-\frac{2\,\Lambda}{3}\,r+\frac{\alpha^2}{r^3}\right)\Bigg].\label{dd2}
\end{eqnarray}

For the Schwarzschild-AdS BH, the spin-dependent potential will be
\begin{eqnarray}
    &&\mathcal{V}_{S, Sch} \simeq \left[1-\frac{2\,M}{r}-\frac{\Lambda}{3}\,r^2\right]\left[\frac{\ell\,(\ell+1)}{r^2}+\frac{S\,(S-1)}{r^2}
    \left(-\frac{2\,M}{r}-\frac{\Lambda}{3}\,r^2\right)+\frac{(1-S)}{r}\left(\frac{2\,M}{r^2}-\frac{2\,\Lambda}{3}\,r\right)\right].\label{sch2}
\end{eqnarray}

For $\xi=1$ which corresponds to ordinary global monopole, the RW-potential from Eq. (\ref{dd2}) becomes
\begin{eqnarray}
    &&\mathcal{V}_S (\mbox{ordinary}) \simeq \left[\mathrm{a}-\frac{2\,M}{r}-\frac{\Lambda}{3}\,r^2+\left(\frac{\Lambda}{2} -\frac{1}{2\,r^2}\right)\,\alpha^2\right]\,\Bigg[\frac{\ell\,(\ell+1)}{r^2}+\frac{S\,(S-1)}{r^2}\times\nonumber\\
    &&\left\{\mathrm{a}-\frac{2\,M}{r}-\frac{\Lambda}{3}\,r^2+\left(\frac{\Lambda}{2} -\frac{1}{2\,r^2}\right)\,\alpha^2-1\right\}
    +\frac{(1-S)}{r}\,\left(\frac{2\,M}{r^2}-\frac{2\,\Lambda}{3}\,r+\frac{\alpha^2}{r^3}\right)\Bigg].\label{ddd2}
\end{eqnarray}

For $\xi=-1$ which corresponds to phantom global monopole, the RW-potential from Eq. (\ref{dd2}) becomes
\begin{eqnarray}
    &&\mathcal{V}_S (\mbox{phantom}) \simeq \left[\mathrm{b}-\frac{2\,M}{r}-\frac{\Lambda}{3}\,r^2+\left(\frac{\Lambda}{2} -\frac{1}{2\,r^2}\right)\,\alpha^2\right]\,\Bigg[\frac{\ell\,(\ell+1)}{r^2}+\frac{S\,(S-1)}{r^2}\times\nonumber\\
    &&\left\{\mathrm{b}-\frac{2\,M}{r}-\frac{\Lambda}{3}\,r^2+\left(\frac{\Lambda}{2} -\frac{1}{2\,r^2}\right)\,\alpha^2-1\right\}
    +\frac{(1-S)}{r}\,\left(\frac{2\,M}{r^2}-\frac{2\,\Lambda}{3}\,r+\frac{\alpha^2}{r^3}\right)\Bigg].\label{dddd2}
\end{eqnarray}

The spin-dependent RW potential (\ref{dd2}) is influenced by the energy scale $\eta$ of the symmetry breaking including the phantom ($\xi = -1$) or the ordinary global monopole ($\xi = 1$), and the quantum correction parameter $\alpha$ for AdS background BH. This result get modifications by the aforementioned parameters compared to those obtained for the Schwarzschild-AdS BH background. Additionally, the potential varies with the multipole quantum number $\ell = 0, 1, 2, \dots$, depending on the spin of the fields involved.

From potential expressions (\ref{ddd2}) and (\ref{dddd2}), we see that the spin-dependent RW potential with ordinary global monopole is lesser than that with phantom global monopole since $\mathrm{a} < \mathrm{b}$.

Below, we calculate this RW-potential for fields of various spins and analyze the result.

\begin{center}
    {\bf I. Zero spin scalar fields (Scalar Perturbations)}
\end{center}

For zero spin scalar field, $S=0$, the RW-potential becomes
\begin{eqnarray}
    \mathcal{V}_0 \simeq \left[1-8\,\pi\,\eta^2\,\xi-\frac{2\,M}{r}-\frac{\Lambda}{3}\,r^2+\left(\frac{\Lambda}{2} -\frac{1}{2\,r^2}\right)\,\alpha^2\right]\,\Bigg[\frac{\ell\,(\ell+1)}{r^2}
    +\frac{2\,M}{r^3}-\frac{2\,\Lambda}{3}+\frac{\alpha^2}{r^4}\Bigg].\label{dd3}
\end{eqnarray}
For the Schwarzschild-AdS BH, this potential will be
\begin{eqnarray}
    \mathcal{V}_{0,Sch} \simeq \left[1-\frac{2\,M}{r}-\frac{\Lambda}{3}\,r^2\right]\,\Bigg[\frac{\ell\,(\ell+1)}{r^2}
    +\frac{2\,M}{r^3}-\frac{2\,\Lambda}{3}\Bigg].\label{sch}
\end{eqnarray}

Considering multipole number $\ell=0$ which corresponds to zero spin $s$-wave scalar field, from Eq. (\ref{dd3}), we find
\begin{eqnarray}
    &&\mathcal{V}_0\Big{|}_{\ell=0} \simeq \left[1-8\,\pi\,\eta^2\,\xi-\frac{2\,M}{r}-\frac{\Lambda}{3}\,r^2+\left(\frac{\Lambda}{2} -\frac{1}{2\,r^2}\right)\,\alpha^2\right]\,\left(\frac{2\,M}{r^3}-\frac{2\,\Lambda}{3}+\frac{\alpha^2}{r^4}\right).\label{dd3a}
\end{eqnarray}

Taking zero cosmological constant ($\Lambda=0$), we introduce dimensionless quantity $x=r/M$ and $y=\alpha/M$. We find a dimensionless quantity from Eq. (\ref{dd3}) defined as,
\begin{eqnarray}
    &&\mbox{Ordinary Global Monopole}:\quad M^2\,\mathcal{V}_0 \simeq \left(\mathrm{a}-\frac{2}{x}-\frac{y^2}{2\,x^2}\right)\,\left(\frac{\ell\,(\ell+1)}{x^2}+\frac{2}{x^3}+\frac{y^2}{x^4} \right),\label{dd3b}\\
    &&\mbox{Phantom Global Monopole}:\quad M^2\,\mathcal{V}_0\simeq \left(\mathrm{b}-\frac{2}{x}-\frac{y^2}{2\,x^2}\right)\,\left(\frac{\ell\,(\ell+1)}{x^2}+\frac{2}{x^3}+\frac{y^2}{x^4} \right),\label{dd3c}
\end{eqnarray}
where $\mathrm{a}=(1-8\,\pi\,\eta^2)<1$ and $\mathrm{b}=(1+8\,\pi\,\eta^2)>1$.

From equations (\ref{dd3b}) and (\ref{dd3c}), we see that the dimensionless quantity $M^2\,\mathcal{V}_0\Big{|}_{\ell=0}$ for zero spin scalar field $s$-wave with ordinary global monopole is lesser than that with phantom global monopole.
\begin{center}
\begin{figure}[ht!]
\subfloat[$\alpha=0.1,\Lambda=-0.1,\ell=1$]{\centering{}\includegraphics[scale=0.32]{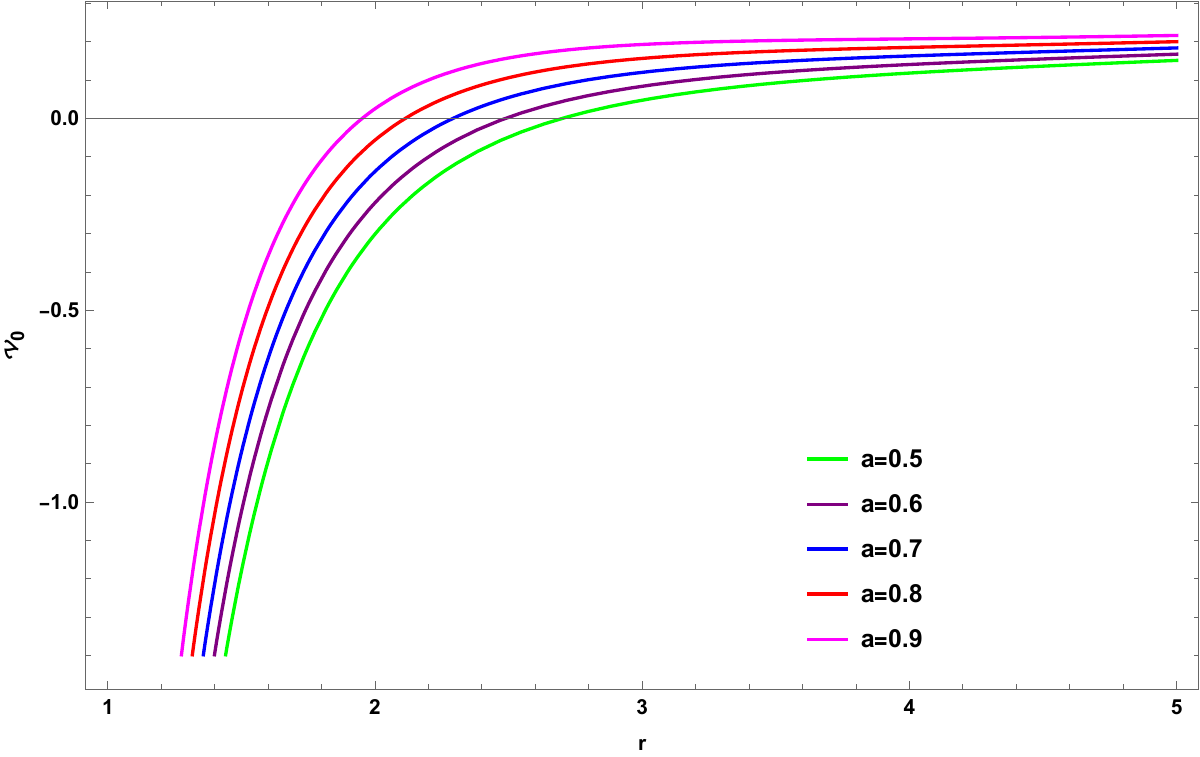}}\quad\quad
\subfloat[$\alpha=0.1,\mathrm{a}=0.9,\ell=1$]{\centering{}\includegraphics[scale=0.32]{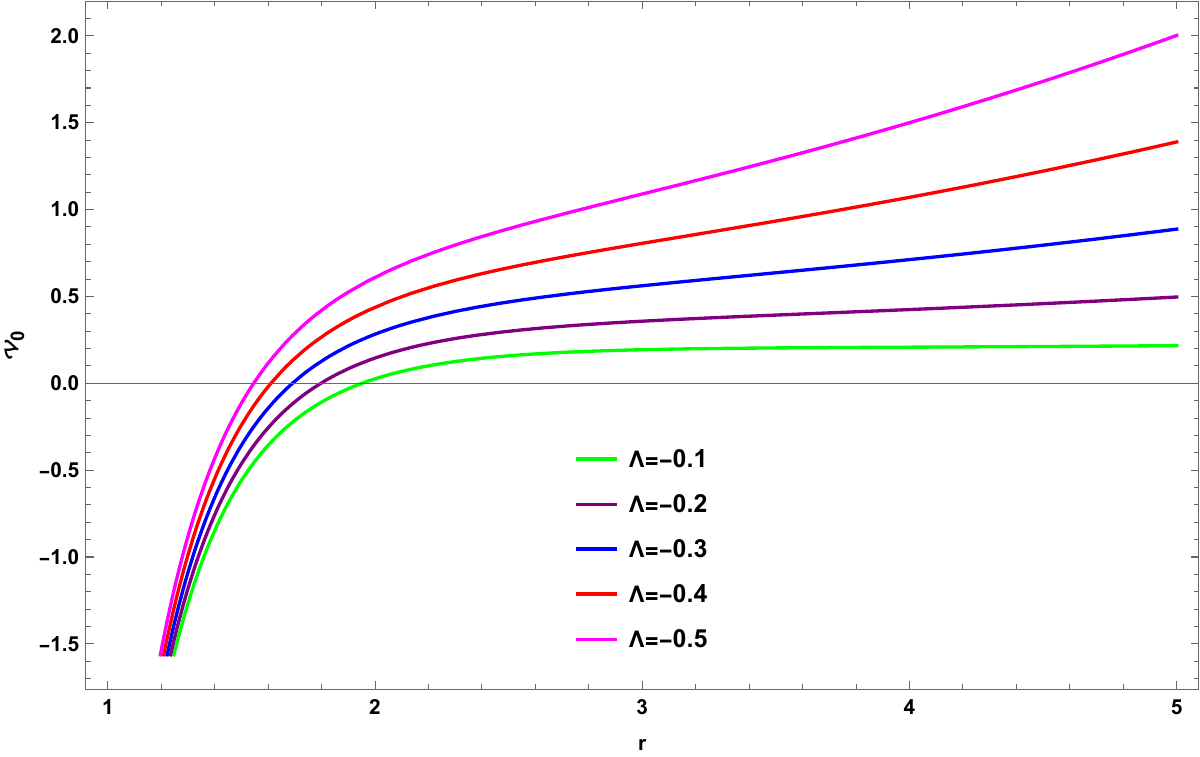}}
\hfill\\
\begin{centering}
\subfloat[$\alpha=0.1,\Lambda=-0.1,\mathrm{a}=0.9$]{\centering{}\includegraphics[scale=0.32]{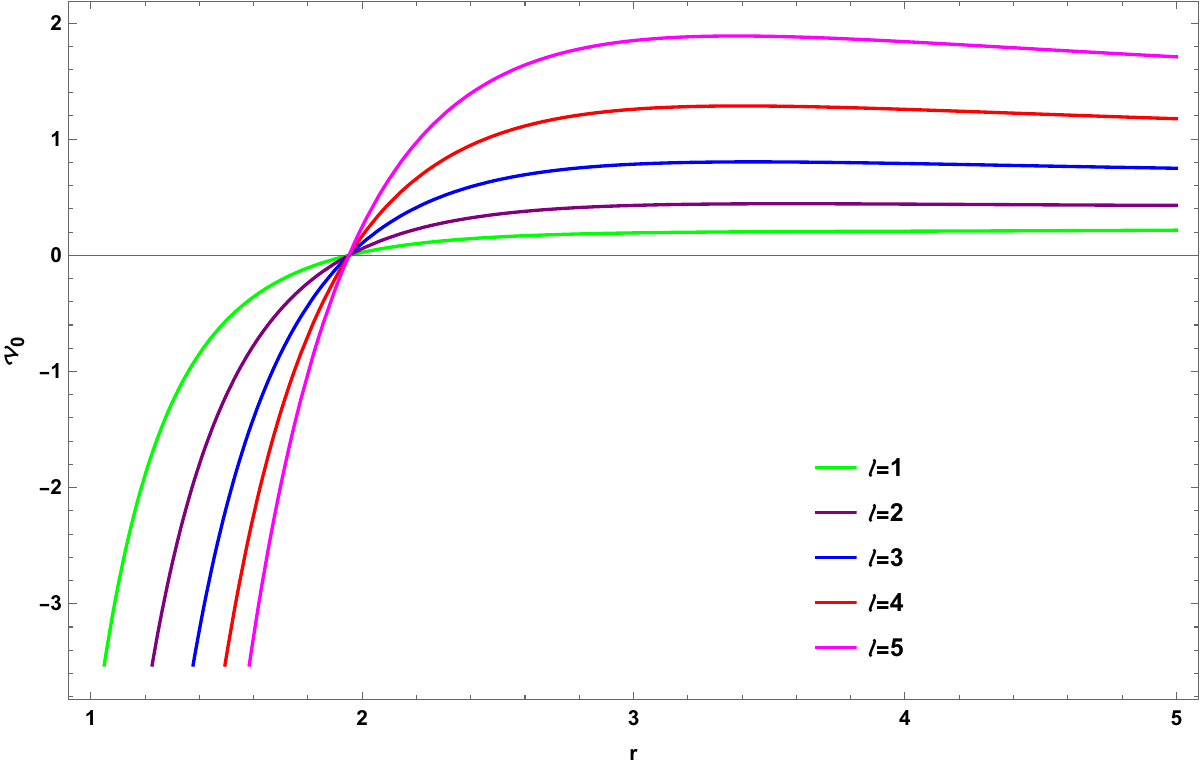}}
\end{centering}
\centering{}\caption{The RW-potential with ordinary global monopole for zero spin scalar field.}\label{fig:1}
\hfill\\
\subfloat[$\alpha=0.1,\Lambda=-0.1,\ell=1$]{\centering{}\includegraphics[scale=0.32]{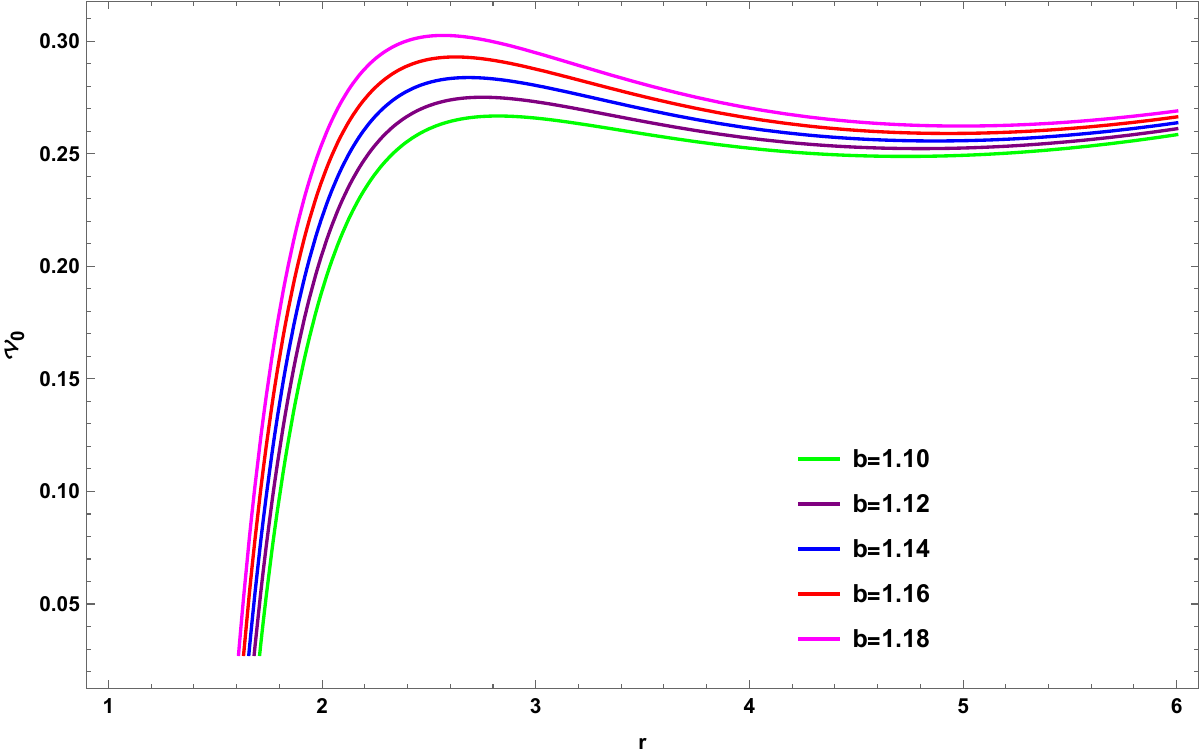}}\quad\quad
\subfloat[$\alpha=0.1,\mathrm{b}=1.1,\ell=1$]{\centering{}\includegraphics[scale=0.32]{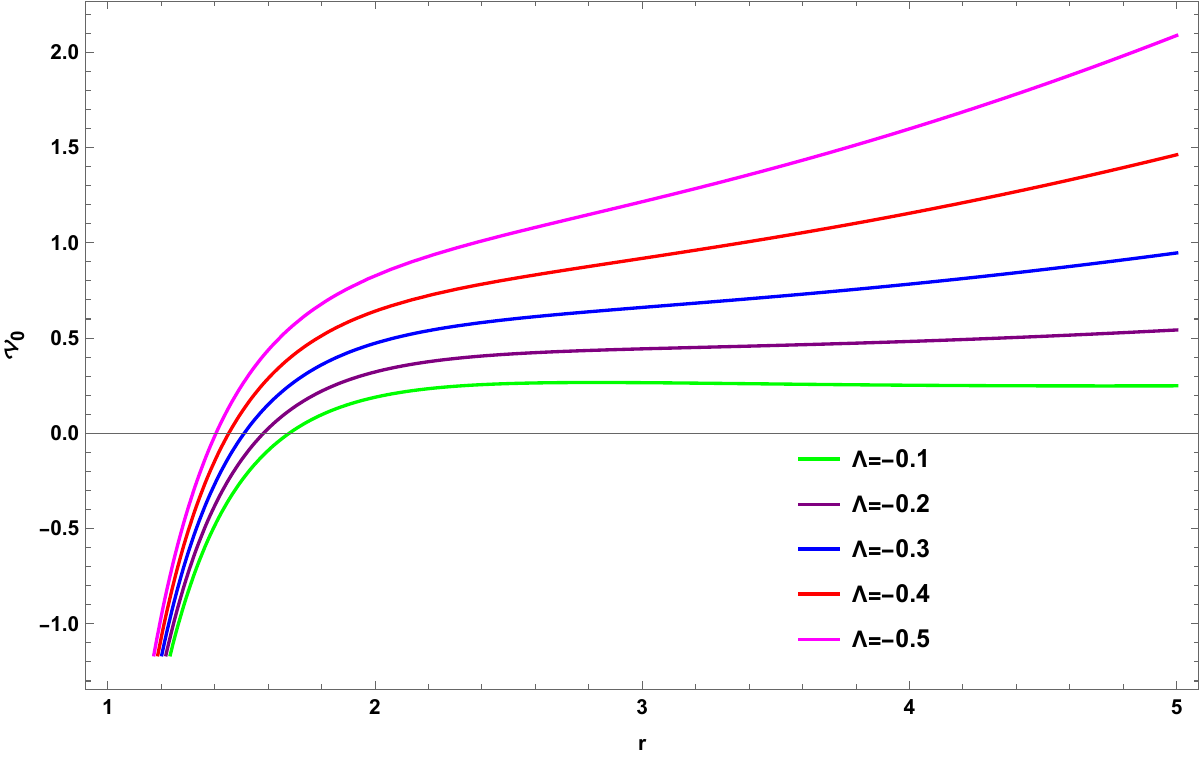}}
\hfill\\
\begin{centering}
\subfloat[$\alpha=0.1,\Lambda=-0.1,\mathrm{b}=1.1$]{\centering{}\includegraphics[scale=0.32]{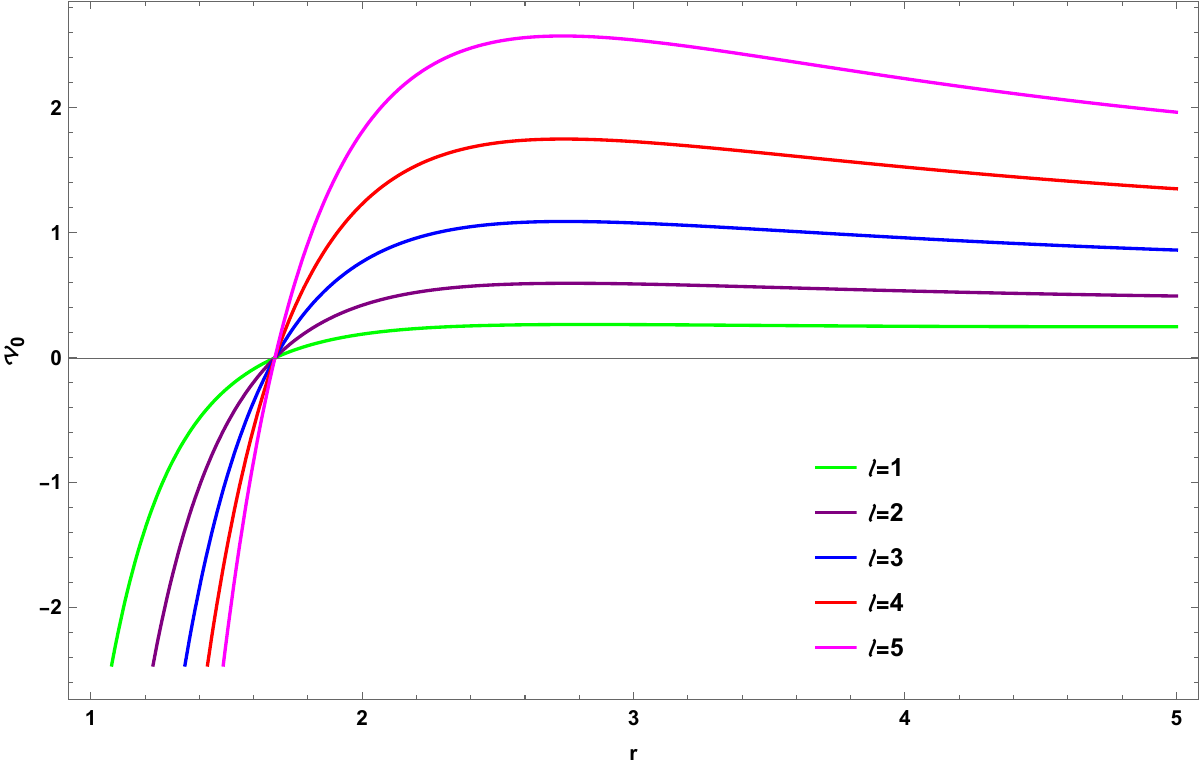}}
\end{centering}
\centering{}\caption{The RW-potential with phantom global monopole for zero spin scalar field.}\label{fig:2}
\end{figure}
\par\end{center}

\begin{center}
\begin{figure}[ht!]
\includegraphics[scale=0.35]{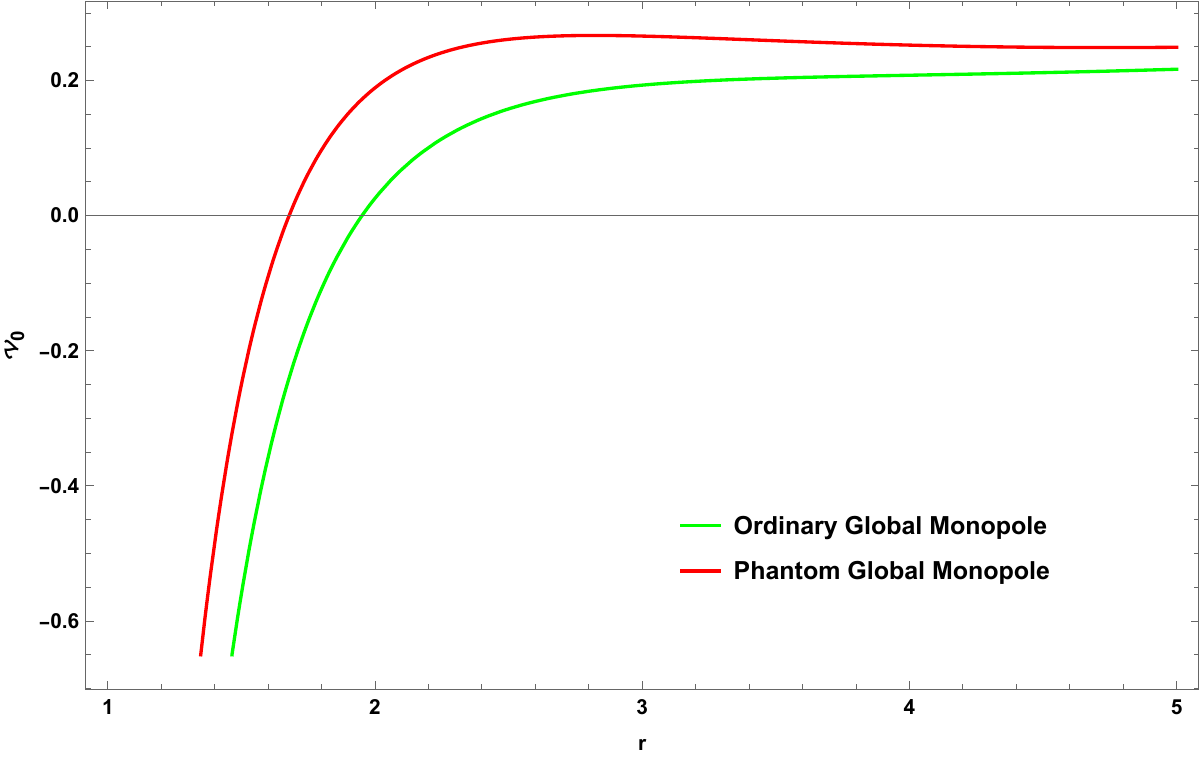}
\centering{}\caption{Comparison of the RW-potential for zero spin scalar field. Here $\alpha=0.1,\Lambda=-0.1,\ell=1,\mathrm{a}=0.9,\mathrm{b}=1.1$.}\label{fig:3}
\end{figure}
\end{center}

\begin{center}
\begin{figure}[ht!]
\includegraphics[scale=0.35]{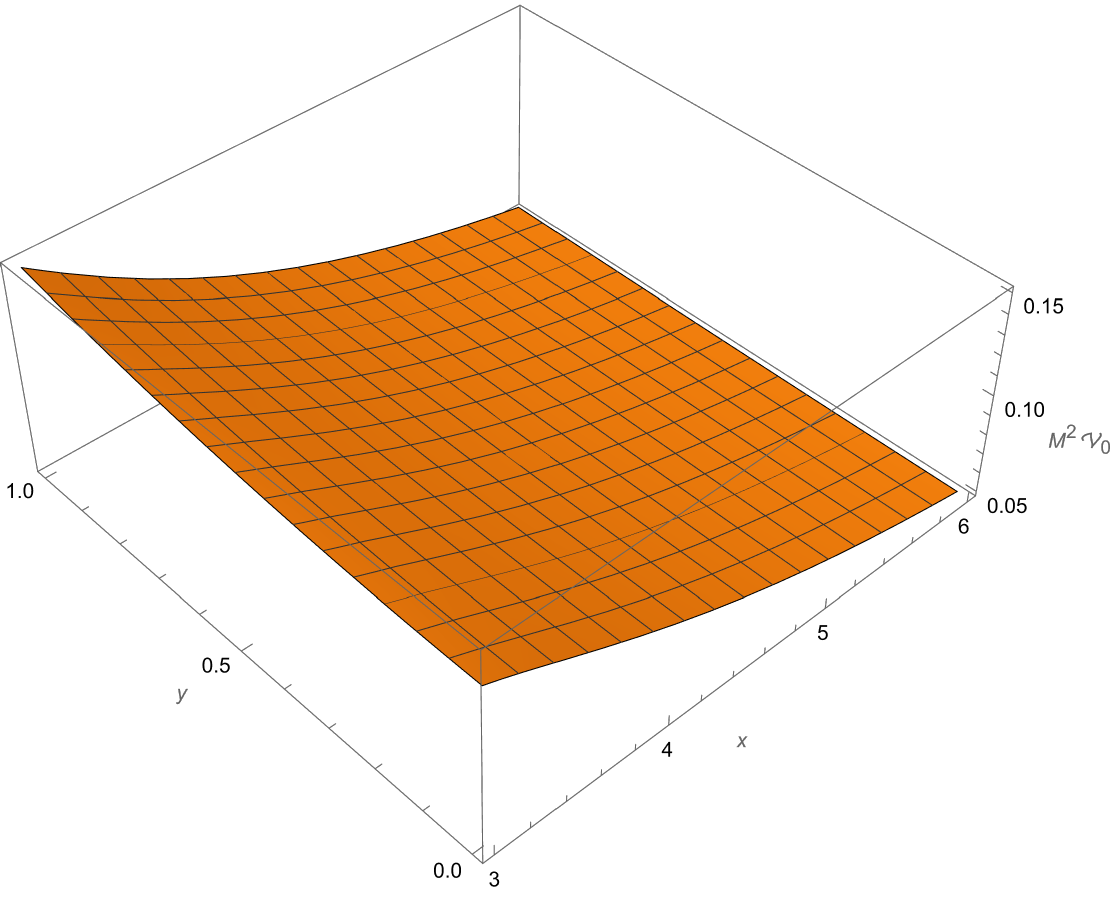}\quad\quad
\includegraphics[scale=0.35]{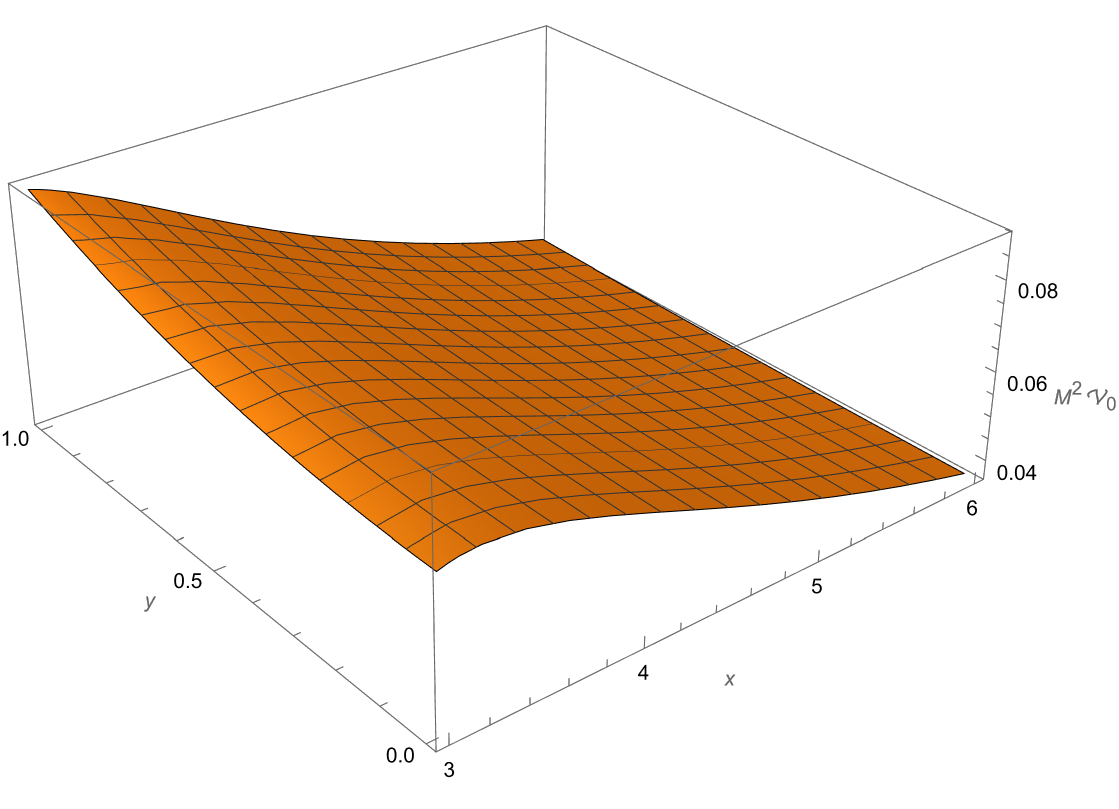}
\centering{}\caption{Graphical representation of the dimensionless quantity (\ref{dd3b}) with multipole number $\ell=0$ (left panel) and $\ell=1$ (right panel). Here $\mathrm{a}=0.9$.}\label{fig:10}
\hfill\\
\includegraphics[scale=0.35]{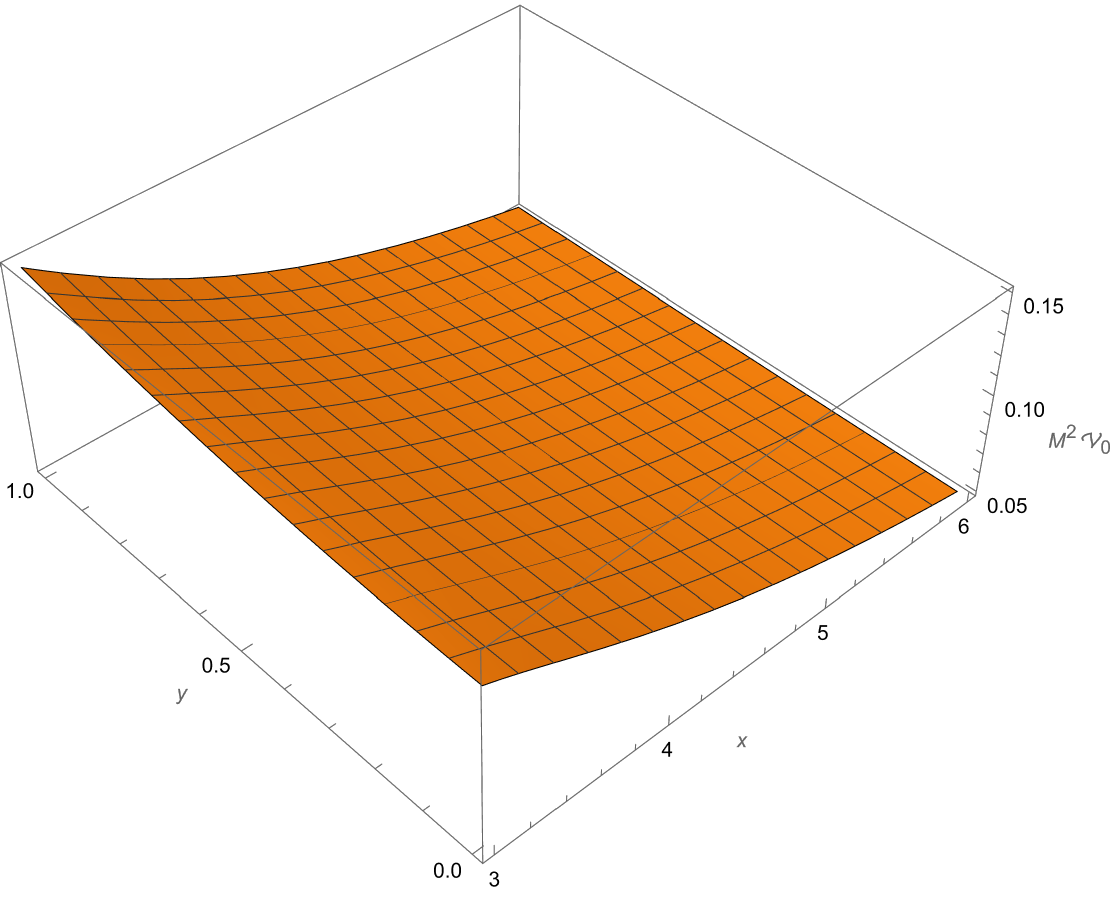}\quad\quad
\includegraphics[scale=0.35]{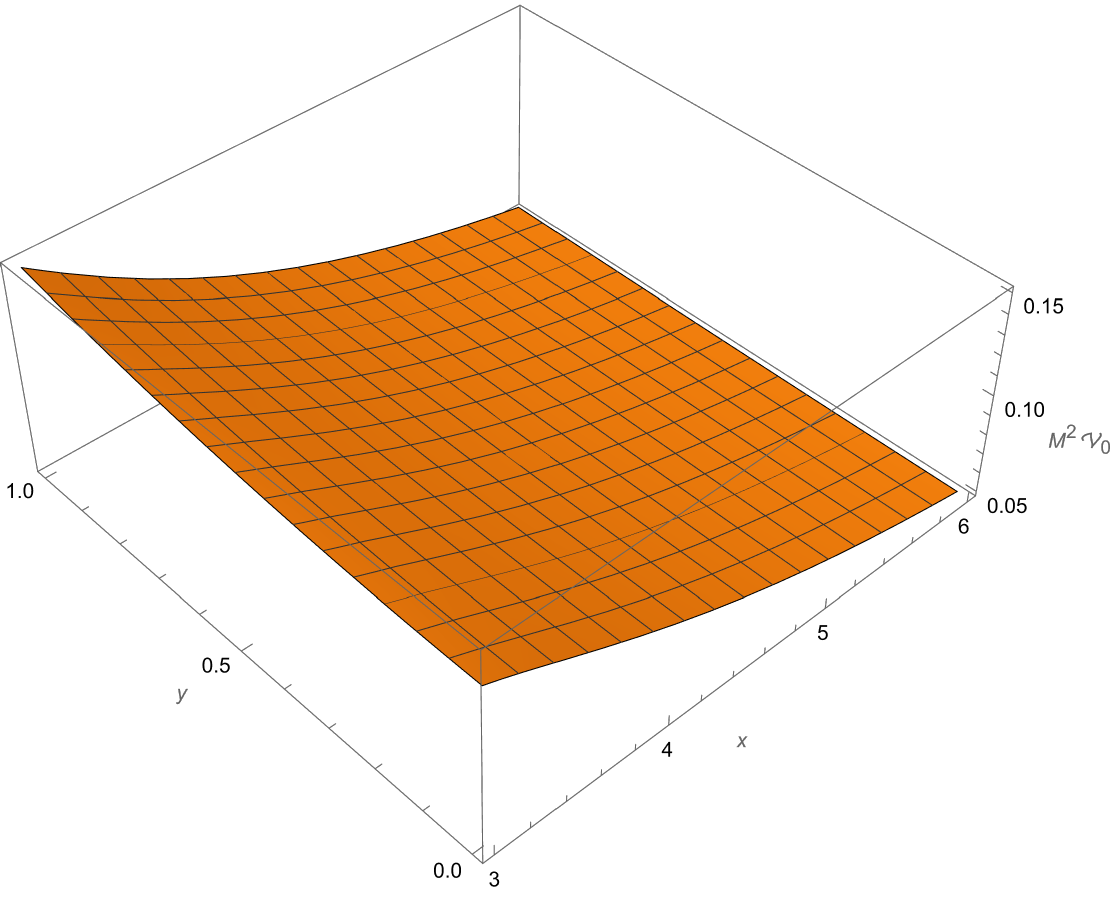}
\centering{}\caption{Graphical representation of the dimensionless quantity (\ref{dd3c}) with multipole number $\ell=0$ (left panel) and $\ell=1$ (right panel). Here $\mathrm{b}=1.1$.}\label{fig:11}
\end{figure}
\end{center}
We have generated Figure \ref{fig:1}, which displays the RW potential for a zero-spin scalar field with an ordinary global monopole, varying different parameters. Similarly, Figure \ref{fig:2} illustrates the RW potential for a zero-spin scalar field with a phantom global monopole, again varying different parameters. In both figures, we observe that changes in various parameters-such as the parameter $\mathrm{a}$, the cosmological constant $\Lambda$, and the multipole number $\ell$-affect the curve behavior. To better understand the differences between the RW potential with an ordinary global monopole and a phantom global monopole, we have generated Figure \ref{fig:3}, which compares the two cases for the multipole number $\ell = 1$.

\begin{center}
\begin{figure}[ht!]
\includegraphics[scale=0.43]{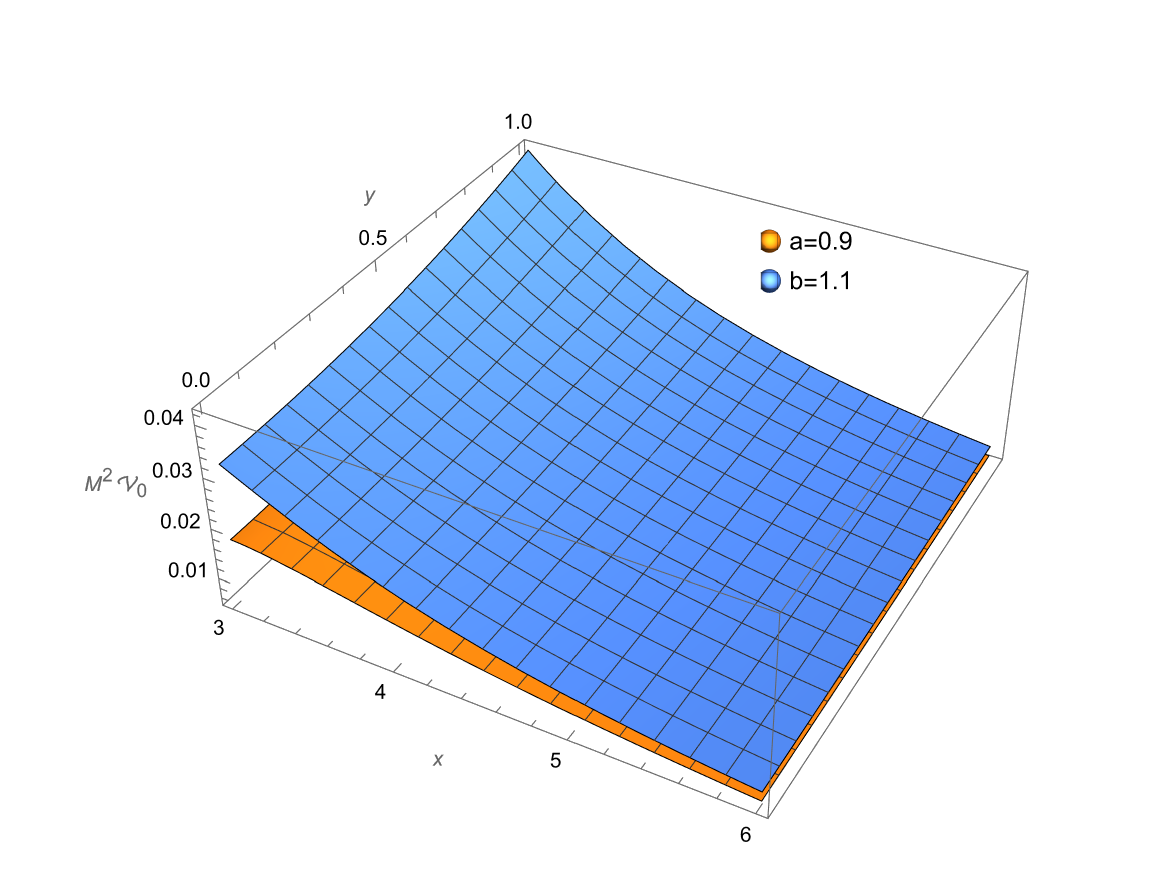}
\includegraphics[scale=0.43]{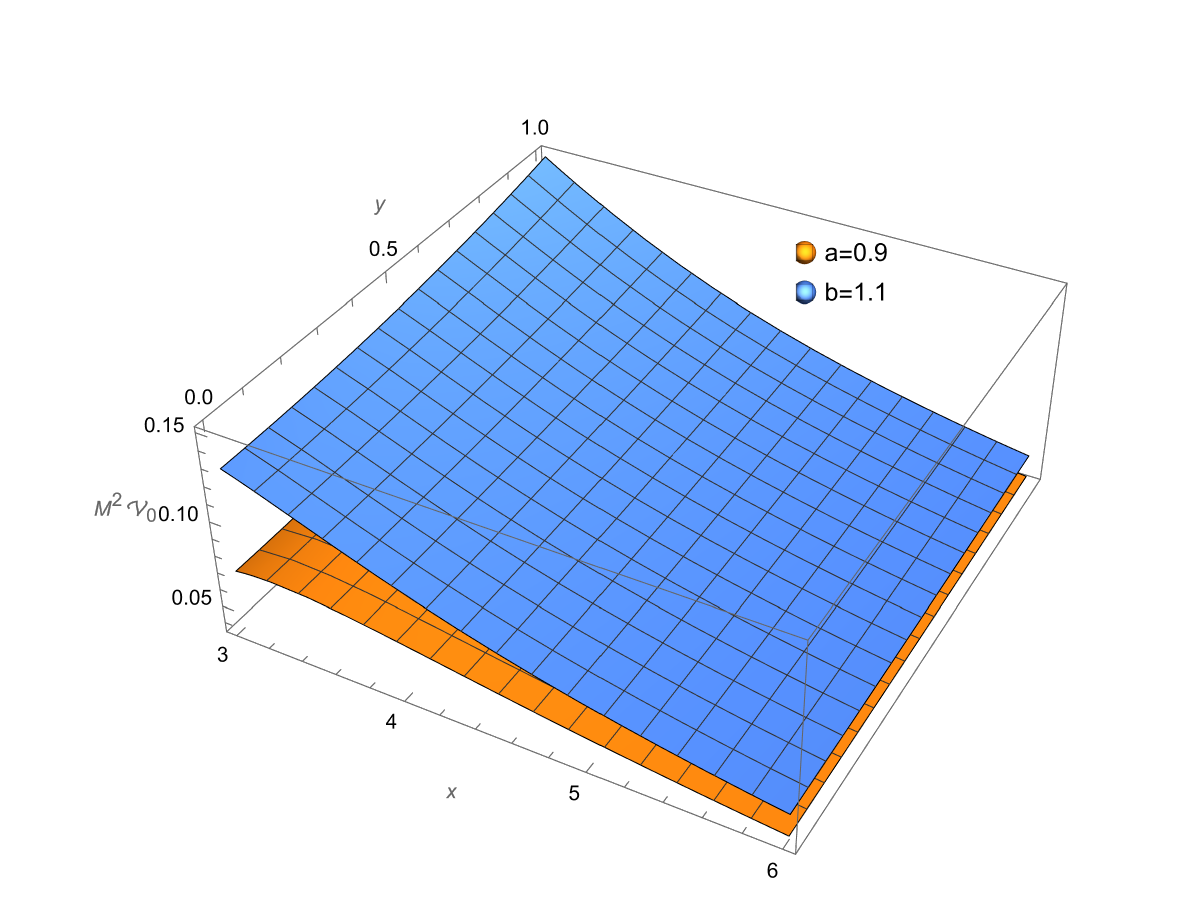}
\centering{}\caption{Comparison of the dimensionless quantity (\ref{dd3b}) and (\ref{dd3c}) with multipole number $\ell=0$ (left panel) and $\ell=1$ (right panel).}\label{fig:12}
\end{figure}
\end{center}

We have generated Figure \ref{fig:10}, which presents a three-dimensional plot of the dimensionless quantity (\ref{dd3b}) with an ordinary global monopole for the multipole numbers $\ell = 0$ (left panel) and $\ell = 1$ (right panel), with the parameter $\mathrm{a} = 0.9$. Similarly, Figure \ref{fig:11} illustrates the dimensionless quantity (\ref{dd3c}) with a phantom global monopole for the same multipole numbers, $\ell = 0$ (left panel) and $\ell = 1$ (right panel), with the parameter $\mathrm{b} = 1.1$. To better understand the differences between the dimensionless quantities (\ref{dd3b}) and (\ref{dd3c}), we have generated Figure \ref{fig:12}, which compares them for the multipole numbers $\ell = 0$ (left panel) and $\ell = 1$ (right panel).
\begin{center}
    {\bf II. Spin-one vector fields: Electromagnetic Perturbations}
\end{center}

For spin-one electromagnetic field, $S=1$, the RW-potential reduces to as follows:
\begin{eqnarray}
    \mathcal{V}_1 \simeq \left[1-8\,\pi\,\eta^2\,\xi-\frac{2\,M}{r}-\frac{\Lambda}{3}\,r^2+\left(\frac{\Lambda}{2} -\frac{1}{2\,r^2}\right)\,\alpha^2\right]\,\frac{\ell\,(\ell+1)}{r^2}
    .\label{dd4}
\end{eqnarray}
For $\xi=1$ which corresponds to ordinary global monopole, the RW-potential for vector fields from Eq. (\ref{dd4}) becomes
\begin{eqnarray}
    \mathcal{V}_1 \simeq \left[\mathrm{a}-\frac{2\,M}{r}-\frac{\Lambda}{3}\,r^2+\left(\frac{\Lambda}{2} -\frac{1}{2\,r^2}\right)\,\alpha^2\right]\,\frac{\ell\,(\ell+1)}{r^2}
    .\label{ddd4}
\end{eqnarray}
For $\xi=-1$ which corresponds to phantom global monopole, the RW-potential for vector fields from Eq. (\ref{dd4}) becomes
\begin{eqnarray}
    \mathcal{V}_1 \simeq \left[\mathrm{b}-\frac{2\,M}{r}-\frac{\Lambda}{3}\,r^2+\left(\frac{\Lambda}{2} -\frac{1}{2\,r^2}\right)\,\alpha^2\right]\,\frac{\ell\,(\ell+1)}{r^2}
    .\label{dddd4}
\end{eqnarray}

From potential expressions (\ref{ddd4}) and (\ref{dddd4}), we see that the RW potential for spin-1 vector fields with ordinary global monopole is lesser than that with phantom global monopole since $\mathrm{a} < \mathrm{b}$. 
\begin{center}
\begin{figure}[ht!]
\subfloat[$\alpha=0.1,\Lambda=-0.1,\ell=1$]{\centering{}\includegraphics[scale=0.32]{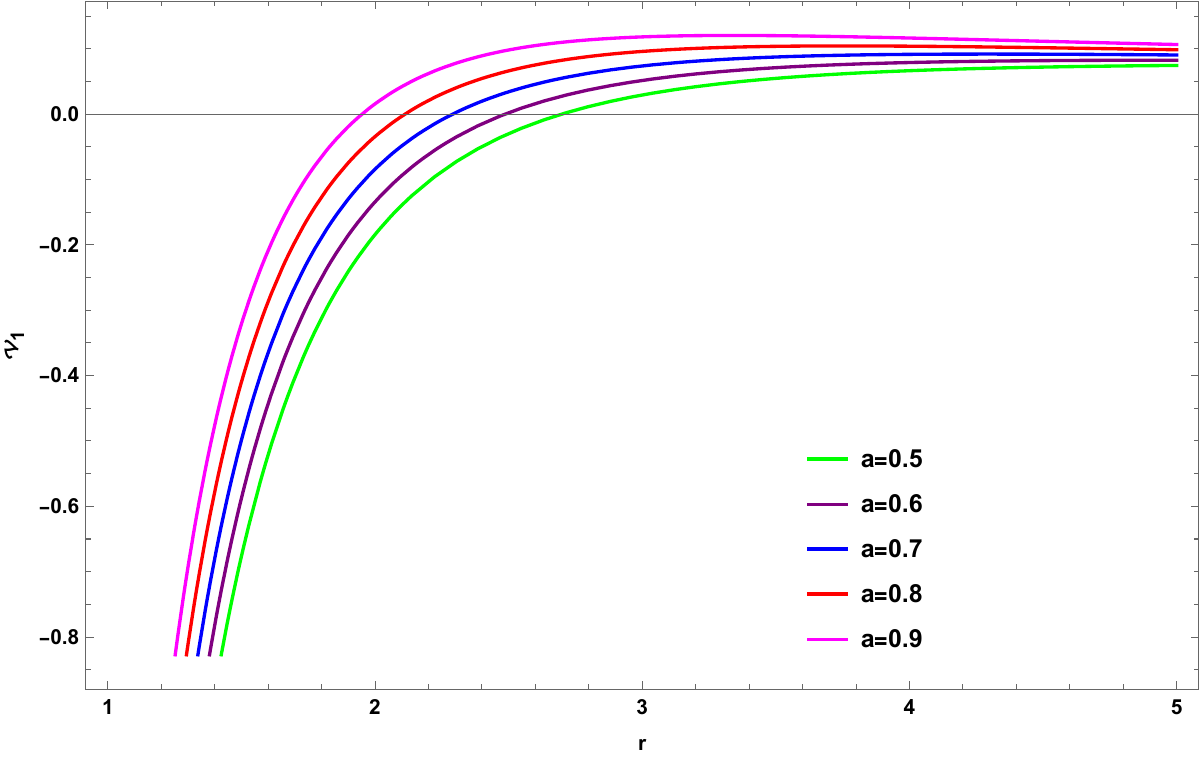}}\quad\quad
\subfloat[$\alpha=0.1,\mathrm{a}=0.9,\ell=1$]{\centering{}\includegraphics[scale=0.32]{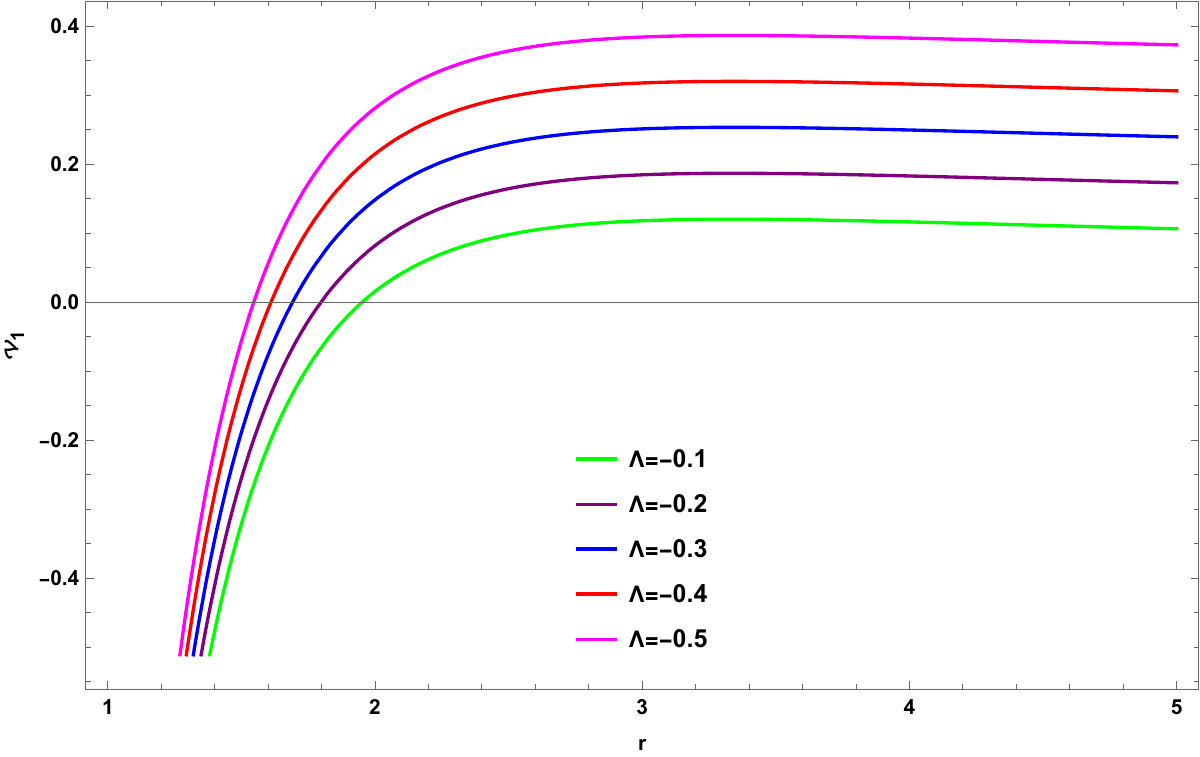}}
\hfill\\
\begin{centering}
\subfloat[$\alpha=0.1,\Lambda=-0.1,\mathrm{a}=0.9$]{\centering{}\includegraphics[scale=0.32]{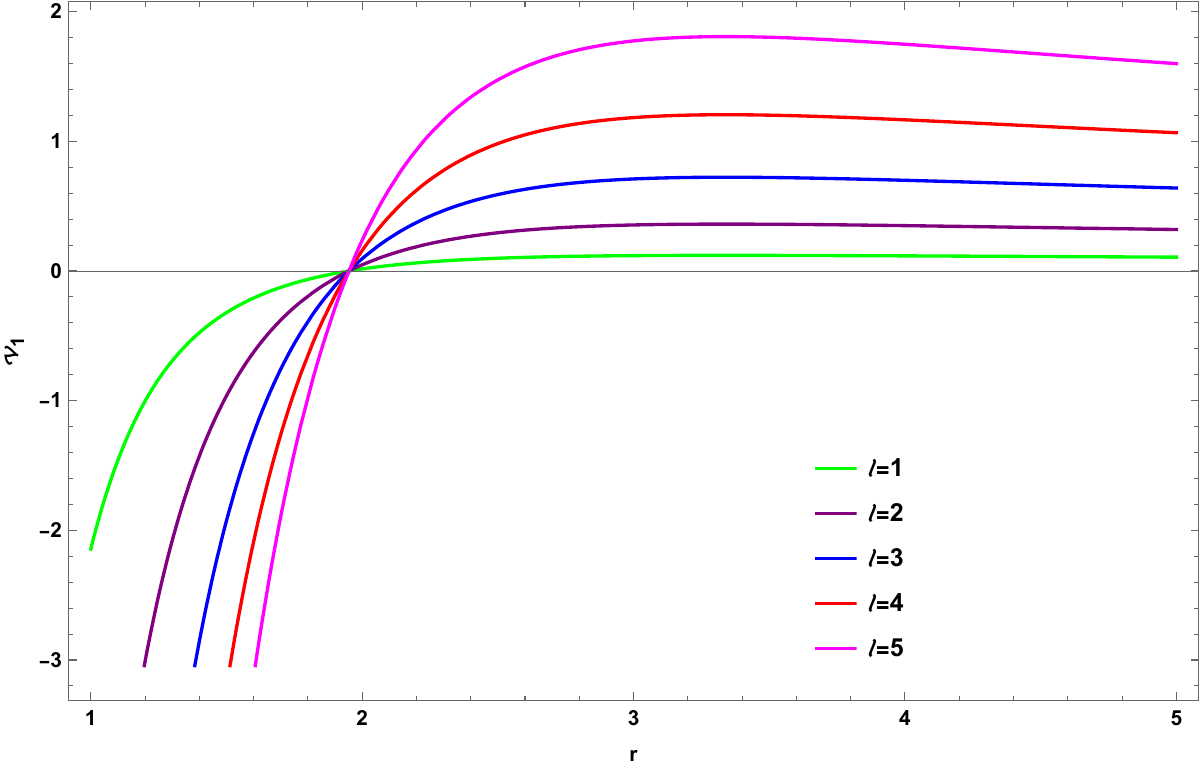}}
\end{centering}
\centering{}\caption{The RW-potential with ordinary global monopole for spin-1 vector field.}\label{fig:4}
\hfill\\
\subfloat[$\alpha=0.1,\Lambda=-0.1,\ell=1$]{\centering{}\includegraphics[scale=0.32]{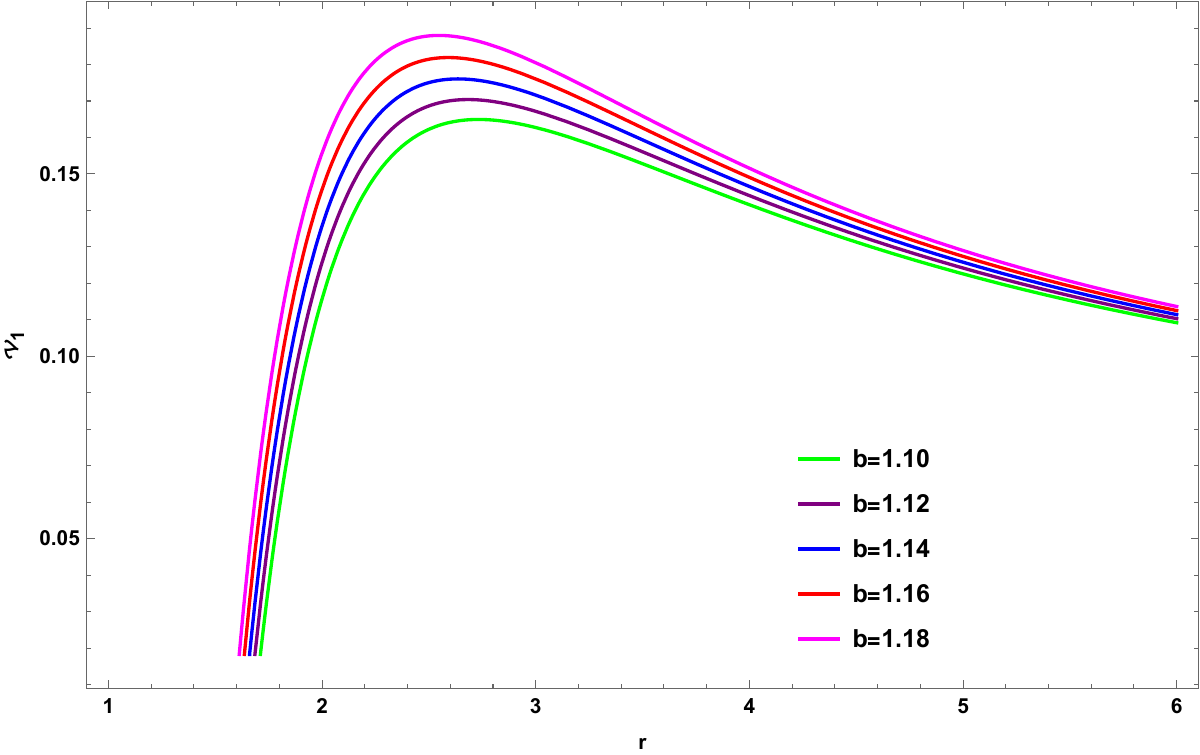}}\quad\quad
\subfloat[$\alpha=0.1,\mathrm{b}=1.1,\ell=1$]{\centering{}\includegraphics[scale=0.32]{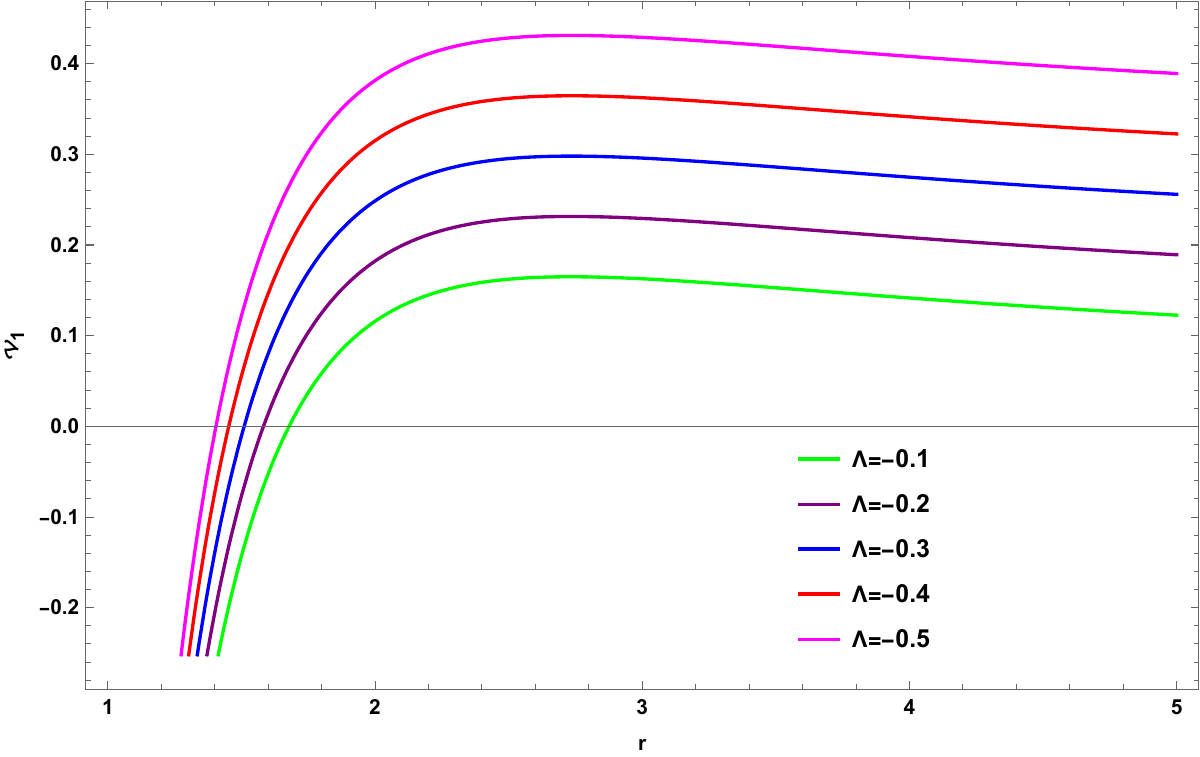}}
\hfill\\
\begin{centering}
\subfloat[$\alpha=0.1,\Lambda=-0.1,\mathrm{b}=1.1$]{\centering{}\includegraphics[scale=0.32]{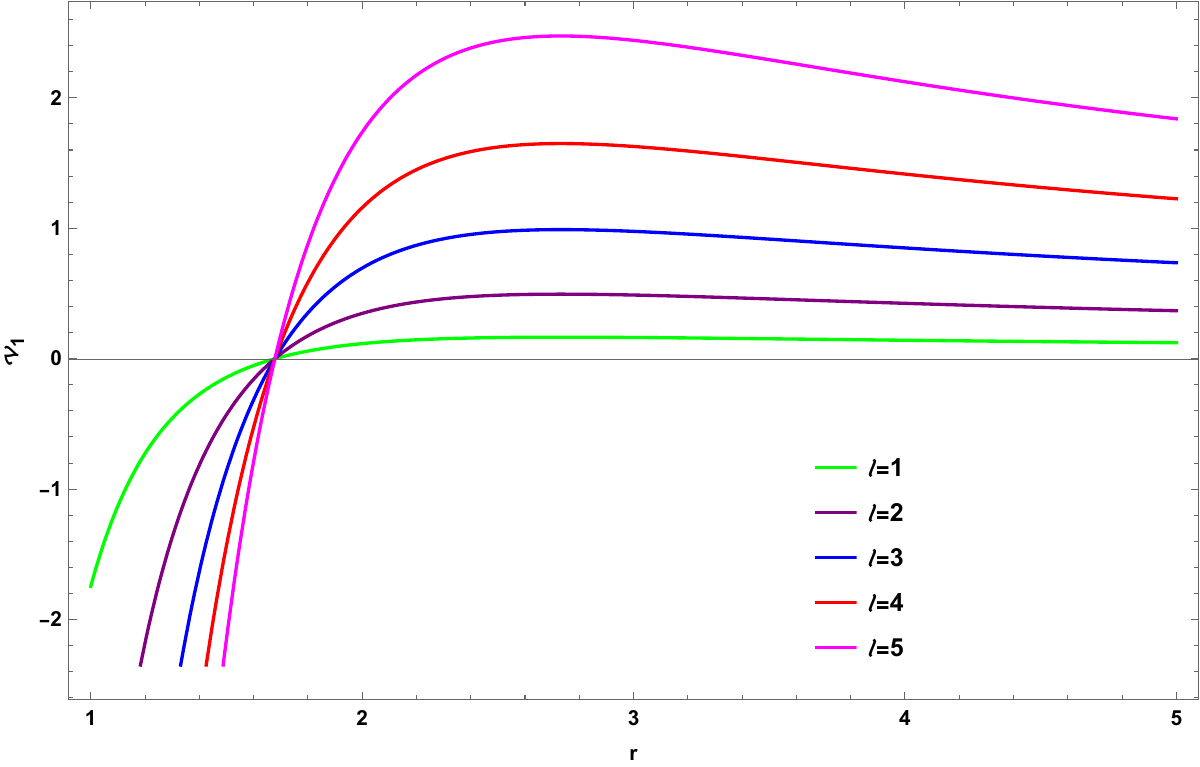}}
\end{centering}
\centering{}\caption{The RW-potential with ordinary global monopole for spin-1 vector field.}\label{fig:5}
\end{figure}
\par\end{center}

\begin{center}
\begin{figure}[ht!]
\includegraphics[scale=0.35]{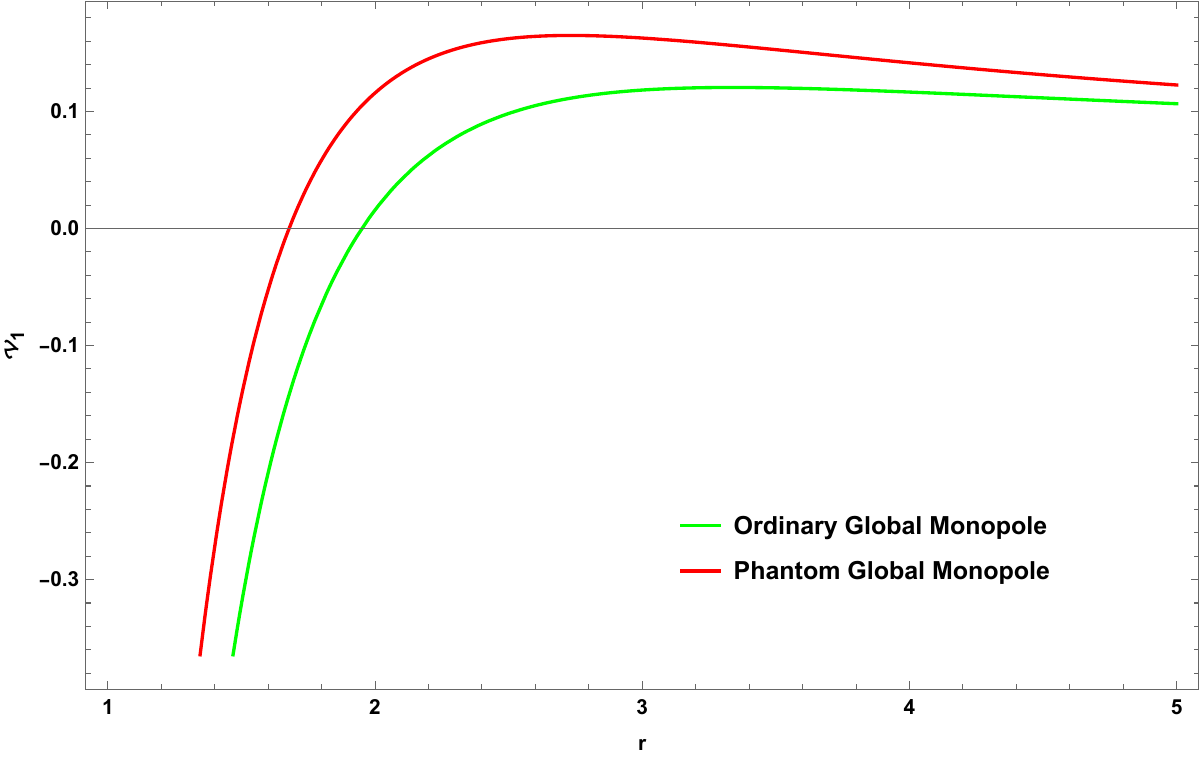}
\centering{}\caption{Comparison of the RW-potential for spin-1 vector field. Here $\alpha=0.1,\Lambda=-0.1,\ell=1,\mathrm{a}=0.9,\mathrm{b}=1.1$.}\label{fig:6}
\end{figure}
\end{center}

We have generated Figure \ref{fig:4}, which displays the RW potential for spin-1 vector fields with an ordinary global monopole, varying different parameters. Similarly, Figure \ref{fig:5} illustrates the RW potential for spin-1 vector fields with a phantom global monopole, again varying different parameters. In both figures, we observe that changes in various parameters-such as the parameter $\mathrm{a}$, the cosmological constant $\Lambda$, and the multipole number $\ell$-affect the curve behavior. To better understand the differences between the RW potential for spin-1 vector fields with an ordinary global monopole and a phantom global monopole, we have generated Figure \ref{fig:6}, which compares the two cases for the multipole number $\ell = 1$.

\begin{center}
    {\bf III. Spin-two tensor fields: Gravitational Perturbations}
\end{center}

For spin-two fields, $S=2$, the RW-potential for spin-2 tensor fields as follows:
\begin{eqnarray}
    &&\mathcal{V}_2 \simeq \left[1-8\,\pi\,\eta^2\,\xi-\frac{2\,M}{r}-\frac{\Lambda}{3}\,r^2+\left(\frac{\Lambda}{2} -\frac{1}{2\,r^2}\right)\,\alpha^2\right]\,\Bigg[\frac{\ell\,(\ell+1)}{r^2}+\frac{2}{r^2}\times\nonumber\\
    &&\left(-8\,\pi\,\eta^2\,\xi-\frac{2\,M}{r}-\frac{\Lambda}{3}\,r^2+\left(\frac{\Lambda}{2}-\frac{1}{2\,r^2}\right)\,\alpha^2\right)
    -\frac{1}{r}\,\left(\frac{2\,M}{r^2}-\frac{2\,\Lambda}{3}\,r+\frac{1}{r^3}\,\alpha^2\right)\Bigg].\label{dd5}
\end{eqnarray} 

\begin{center}
\begin{figure}[ht!]
\subfloat[$\alpha=0.1,\Lambda=-0.1,\ell=1$]{\centering{}\includegraphics[scale=0.32]{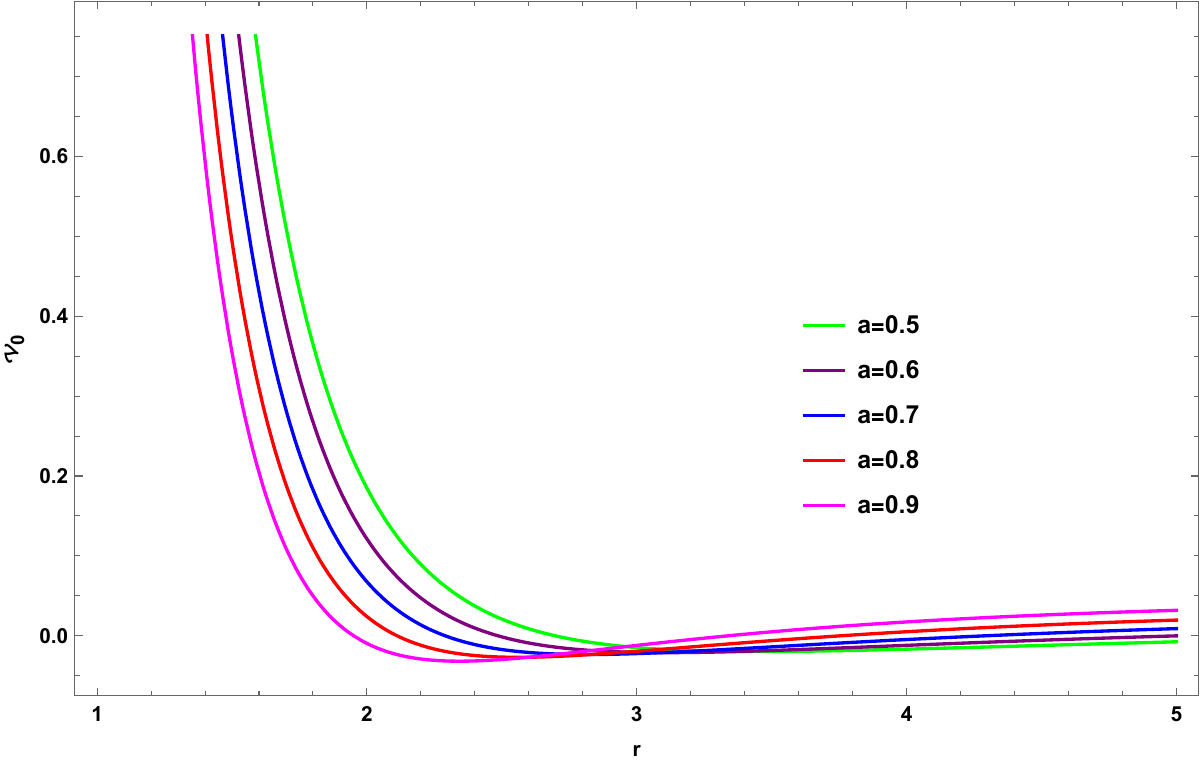}}\quad\quad
\subfloat[$\alpha=0.1,\mathrm{a}=0.9,\ell=1$]{\centering{}\includegraphics[scale=0.32]{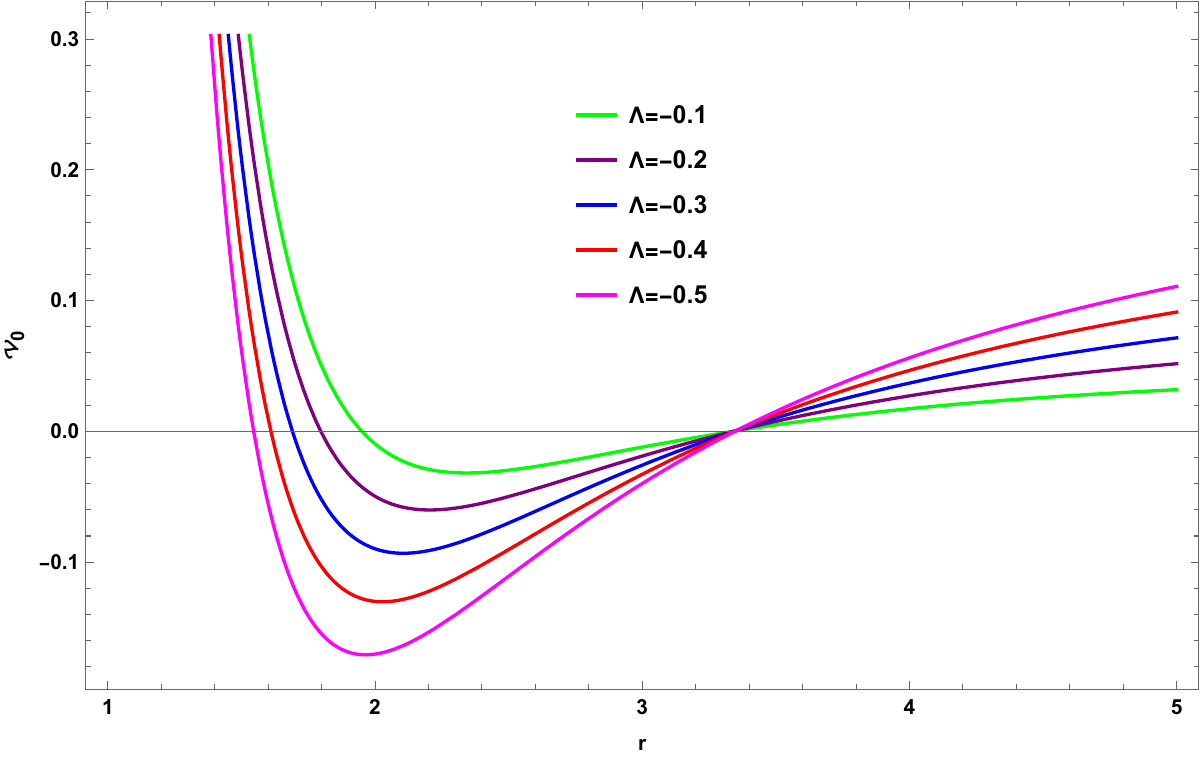}}
\hfill\\
\begin{centering}
\subfloat[$\alpha=0.1,\Lambda=-0.1,\mathrm{a}=0.9$]{\centering{}\includegraphics[scale=0.32]{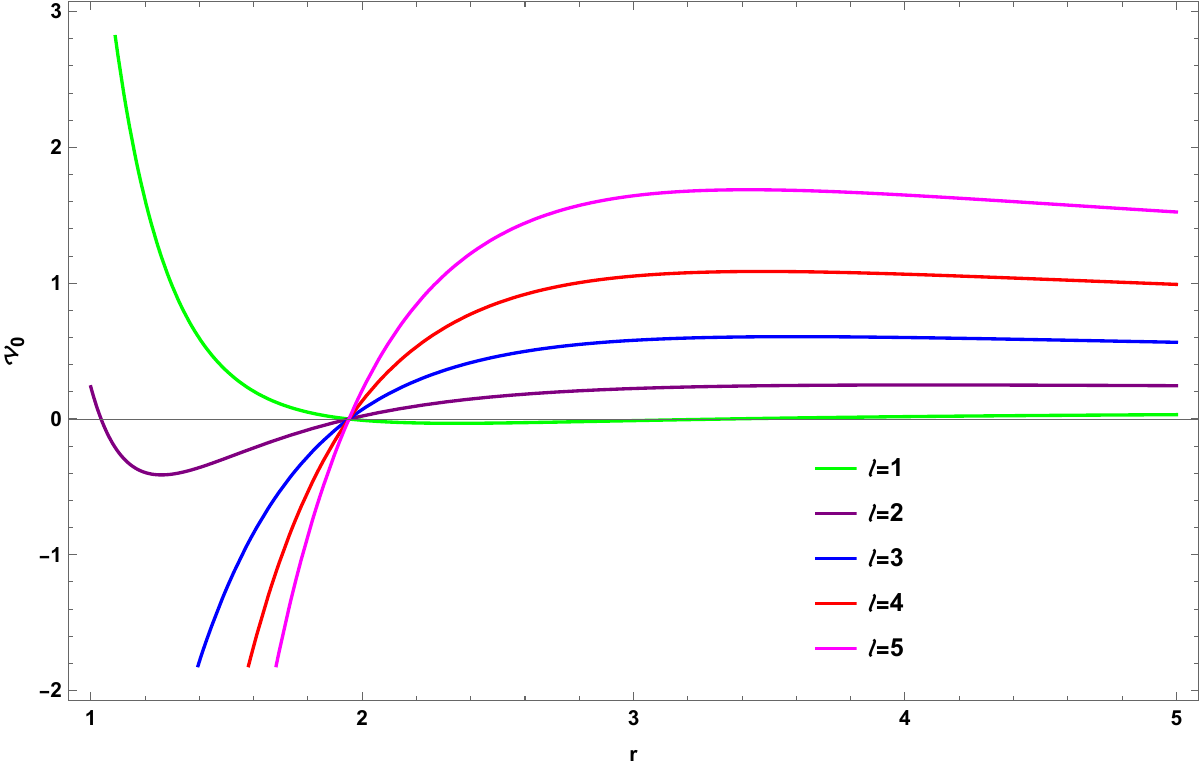}}
\end{centering}
\centering{}\caption{The RW-potential with ordinary global monopole for spin-2 tensor field.}\label{fig:7}
\end{figure}
\par\end{center}

\begin{center}
\begin{figure}[ht!]
\subfloat[$\alpha=0.1,\Lambda=-0.1,\ell=1$]{\centering{}\includegraphics[scale=0.32]{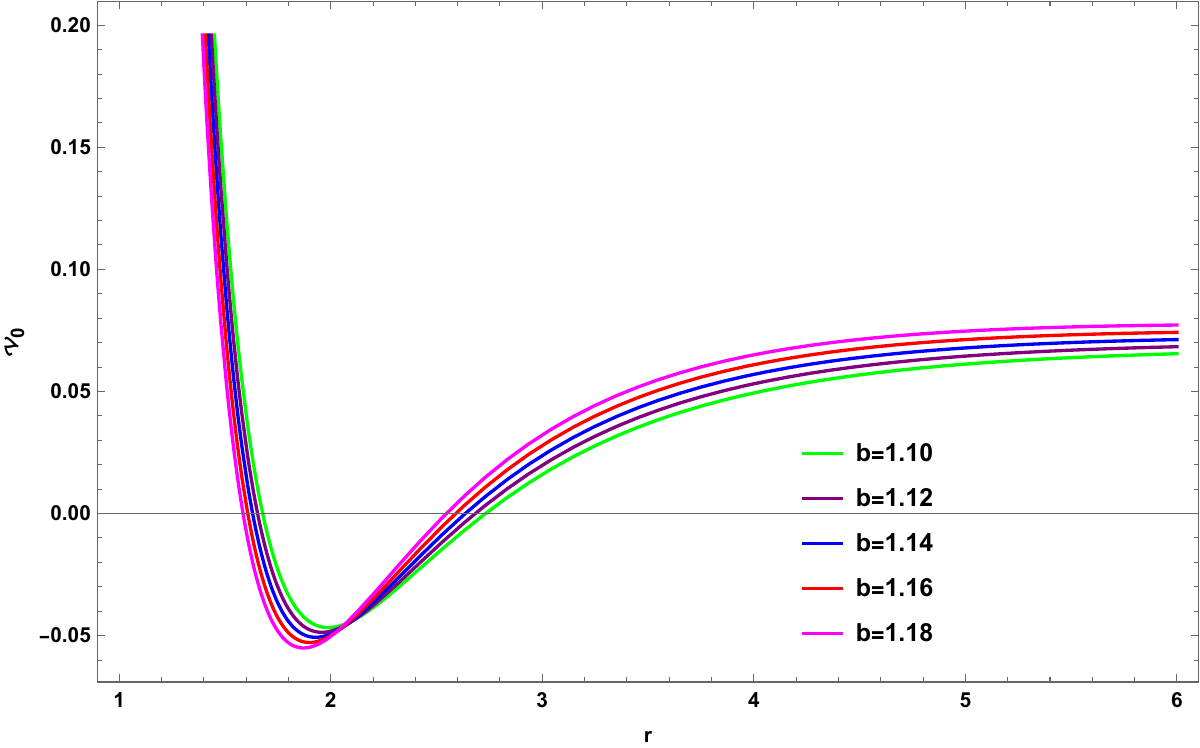}}\quad\quad
\subfloat[$\alpha=0.1,\mathrm{b}=1.1,\ell=1$]{\centering{}\includegraphics[scale=0.32]{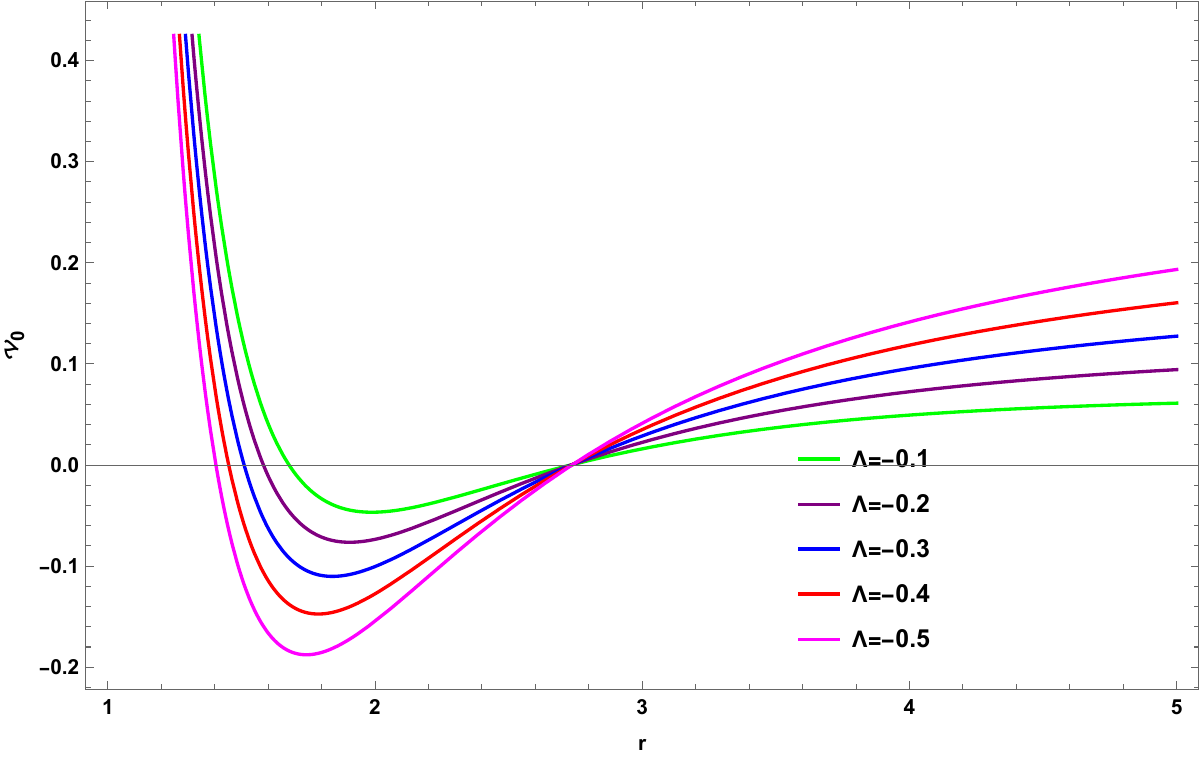}}
\hfill\\
\begin{centering}
\subfloat[$\alpha=0.1,\Lambda=-0.1,\mathrm{b}=1.1$]{\centering{}\includegraphics[scale=0.32]{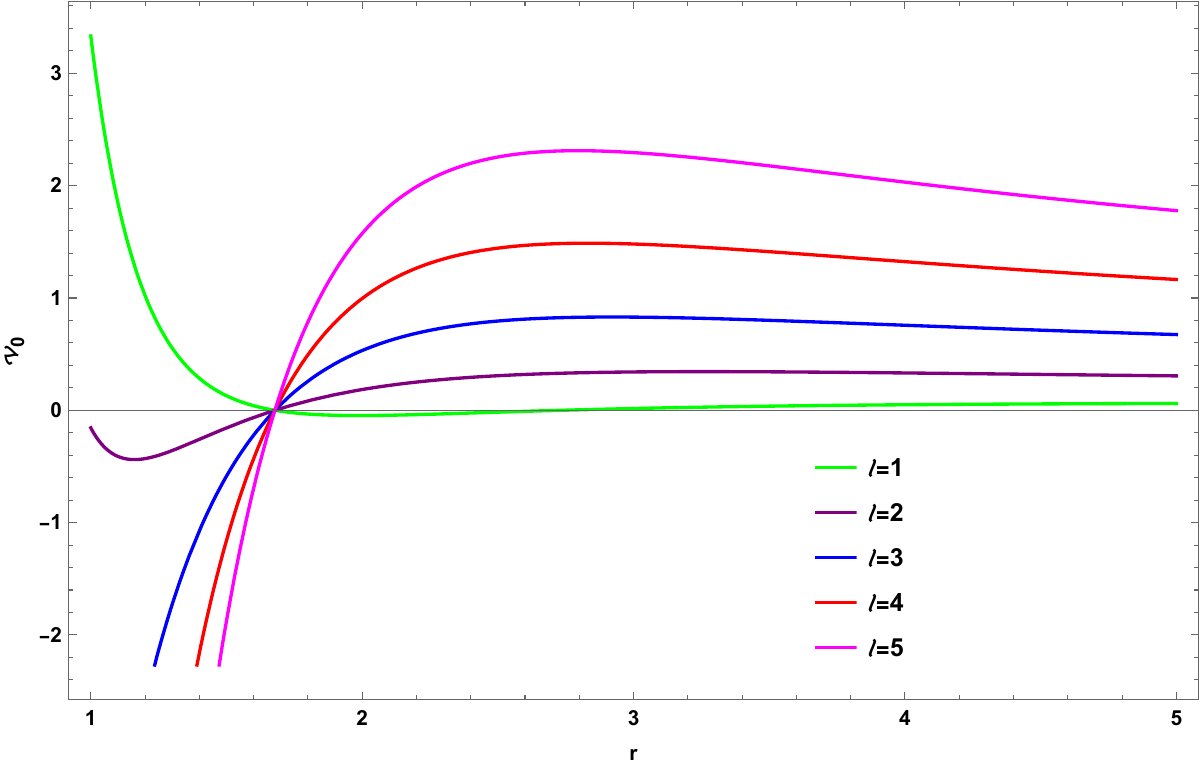}}
\end{centering}
\centering{}\caption{The RW-potential with phantom global monopole for spin-2 tensor field.}\label{fig:8}
\end{figure}
\par\end{center}

\begin{center}
\begin{figure}[ht!]
\includegraphics[scale=0.35]{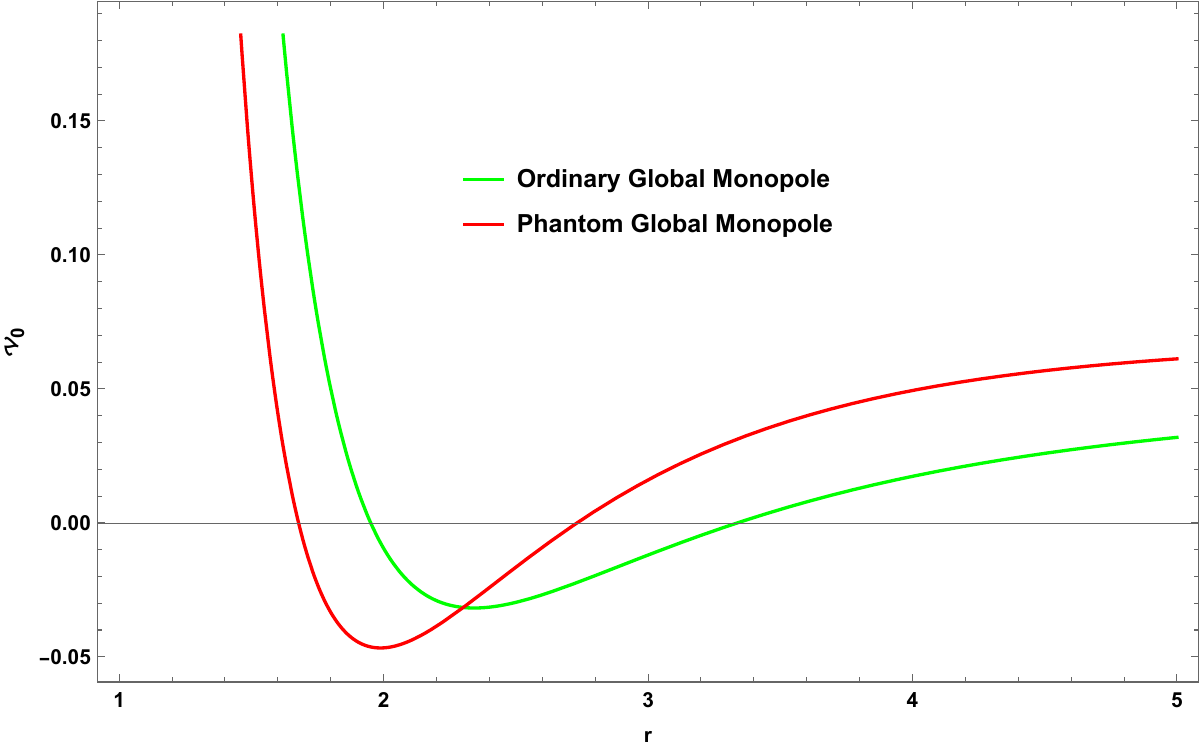}
\centering{}\caption{Comparison of the RW-potential for spin-2 tensor field. Here $\alpha=0.1,\Lambda=-0.1,\ell=1,\mathrm{a}=0.9,\mathrm{b}=1.1$.}\label{fig:9}
\end{figure}
\end{center}

For $\xi=1$ which corresponds to ordinary global monopole, the RW-potential for spin-2 tensor fields from Eq. (\ref{dd5}) becomes
\begin{eqnarray}
    &&\mathcal{V}_2 \simeq \left[\mathrm{a}-\frac{2\,M}{r}-\frac{\Lambda}{3}\,r^2+\left(\frac{\Lambda}{2} -\frac{1}{2\,r^2}\right)\,\alpha^2\right]\,\Bigg[\frac{\ell\,(\ell+1)}{r^2}+\frac{2}{r^2}\times\nonumber\\
    &&\left\{\mathrm{a}-\frac{2\,M}{r}-\frac{\Lambda}{3}\,r^2+\left(\frac{\Lambda}{2}-\frac{1}{2\,r^2}\right)\,\alpha^2-1\right\}
    -\frac{1}{r}\,\left(\frac{2\,M}{r^2}-\frac{2\,\Lambda}{3}\,r+\frac{1}{r^3}\,\alpha^2\right)\Bigg].\label{dd6}
\end{eqnarray}

For $\xi=-1$ which corresponds to phantom global monopole, the RW-potential for spin-2 tensor fields from Eq. (\ref{dd5}) becomes
\begin{eqnarray}
    &&\mathcal{V}_2 \simeq \left[\mathrm{b}-\frac{2\,M}{r}-\frac{\Lambda}{3}\,r^2+\left(\frac{\Lambda}{2} -\frac{1}{2\,r^2}\right)\,\alpha^2\right]\,\Bigg[\frac{\ell\,(\ell+1)}{r^2}+\frac{2}{r^2}\times\nonumber\\
    &&\left\{\mathrm{b}-\frac{2\,M}{r}-\frac{\Lambda}{3}\,r^2+\left(\frac{\Lambda}{2}-\frac{1}{2\,r^2}\right)\,\alpha^2-1\right\}
    -\frac{1}{r}\,\left(\frac{2\,M}{r^2}-\frac{2\,\Lambda}{3}\,r+\frac{1}{r^3}\,\alpha^2\right)\Bigg].\label{dd7}
\end{eqnarray}

From potential expressions (\ref{dd6}) and (\ref{dd7}), we see that the RW potential for spin-2 tensor fields with ordinary global monopole is lesser than that with phantom global monopole since $\mathrm{a} < \mathrm{b}$. 

We have generated Figure \ref{fig:7}, which displays the RW potential for spin-2 tensor fields with an ordinary global monopole, varying different parameters. Similarly, Figure \ref{fig:8} illustrates the RW potential for spin-2 tensor fields with a phantom global monopole, again varying different parameters. In both figures, we observe that changes in various parameters-such as the parameter $\mathrm{a}$, the cosmological constant $\Lambda$, and the multipole number $\ell$-affect the curve behavior. To better understand the differences between the RW potential for spin-2 tensor fields with an ordinary global monopole and a phantom global monopole, we have generated Figure \ref{fig:9}, which compares the two cases for the multipole number $\ell = 1$.

\subsection{QNM Frequency}

This part aims is to calculate the  quasinormal frequencies of scalar field in Eq. (\ref{dd3}) and the electromagnetic field in Eq. (\ref{dd4}) with zero cosmological constant ($\Lambda=0$). We will use the 6th-order WKB approximation \cite{Iyer, Konoplya} with a bigger value of $l$. The reason for choosing a somewhat higher value of multipole moment $l$ is that the error associated with the WKB technique decreases significantly as $l$ increases. Tables \ref{taba1} and \ref{taba2} summaries the findings.  

\begin{center}
\begin{tabular}{|c|c|c|c|}
 \hline 
   \multicolumn{4}{|c|}{ EM ($l=1$, $n=0$, $M=1$, $\eta=0.1$)}
\\ \hline 
 $\alpha$ & $\xi =-1$ & $0$ & $1$ \\ \hline
$0.0$ & $0.33889-0.143706i$ & $0.24819-0.092637i$ & $0.164697-0.052418i$ \\ 
$0.04$ & $0.33882-0.143695i$ & $0.24815-0.092632i$ & $0.164679-0.052416i$ \\ 
$0.08$ & $0.33862-0.143663i$ & $0.24804-0.092616i$ & $0.164624-0.052410i$ \\ 
$0.12$ & $0.33829-0.14361i$ & $0.24785-0.09259i$ & $0.164534-0.052400i$ \\ 
$0.16$ & $0.33782-0.143536i$ & $0.24759-0.092555i$ & $0.164407-0.052386i$ \\ 
$0.2$ & $0.33723-0.143441i$ & $0.24725-0.09250i$ & $0.164246-0.052368i$\\  \hline \multicolumn{4}{|c|}{  ($l=2$)}
\\ \hline 
$0.0$ & $0.63566-0.148291i$ & $0.45759-0.095011i$ & $0.298675-0.053425i$ \\ 
$0.04$ & $0.63555-0.14828i$ & $0.45753-0.095006i$ & $0.298644-0.053423i$ \\ 
$0.08$ & $0.63521-0.14826i$ & $0.457339-0.094993i$ & $0.298552-0.053417i$ \\ 
$0.12$ & $0.63466-0.14821i$ & $0.457022-0.094970i$ & $0.298398-0.053408i$ \\ 
$0.16$ & $0.63388-0.148153i$ & $0.456579-0.094939i$ & $0.298184-0.053395i$ \\ 
$0.2$ & $0.63289-0.148073i$ & $0.456013-0.094899i$ & $0.29791-0.053378i$%
\\ 
 \hline
\end{tabular}
\captionof{table}{The QNM of the quantum corrected AdS BH with Phantom Global Monopoles  for EM perturbation.} \label{taba1}
\end{center}

\begin{center}
\begin{tabular}{|c|c|c|c|}
 \hline 
   \multicolumn{4}{|c|}{Scalar ($l=1$, $n=0$, $M=1$, $\eta=0.1$) }
\\ \hline 
$\alpha$ & $\xi =-1$ & $0$ & $1$ \\ \hline
$0.0$ & $0.41703-0.153588i$ & $0.29291-0.097761i$ & $0.18644-0.054594i$ \\ 
$0.04$ & $0.41696-0.153582i$ & $0.29287-0.097757i$ & $0.18643-0.054593i$ \\ 
$0.08$ & $0.41675-0.153559i$ & $0.29275-0.097745i$ & $0.18637-0.054588i$ \\ 
$0.12$ & $0.41640-0.153520i$ & $0.29255-0.097725i$ & $0.18627-0.054579i$ \\ 
$0.16$ & $0.41591-0.153467i$ & $0.29227-0.097697i$ & $0.18614-0.054567i$ \\ 
$0.2$ & $0.41528-0.153397i$ & $0.29192-0.097661i$ & $0.18598-0.054551i$ \\  
\hline          
\multicolumn{4}{|c|}{ ($l=2$) }\\ \hline
$0.0$ & $0.681234-0.151723i$ & $0.483642-0.096766i$ & $0.31132-0.054164i$ \\ 
$0.04$ & $0.681119-0.151715i$ & $0.483577-0.096761i$ & $0.31129-0.054162i$ \\ 
$0.08$ & $0.680777-0.151691i$ & $0.483383-0.096749i$ & $0.31120-0.054157i$ \\ 
$0.12$ & $0.680208-0.15165i$ & $0.48306-0.096728i$ & $0.31104-0.054148i$ \\ 
$0.16$ & $0.679415-0.15159i$ & $0.48261-0.096698i$ & $0.31082-0.054135i$ \\ 
$0.2$ & $0.678401-0.15151i$ & $0.482033-0.096660i$ & $0.31054-0.054119i$
\\ 
 \hline
\end{tabular}
\captionof{table}{The QNM of the quantum corrected AdS BH with Phantom Global Monopoles  for scalar perturbation.} \label{taba2}
\end{center}

\begin{figure}[ht!]
    \centering
    \includegraphics[scale=0.7]{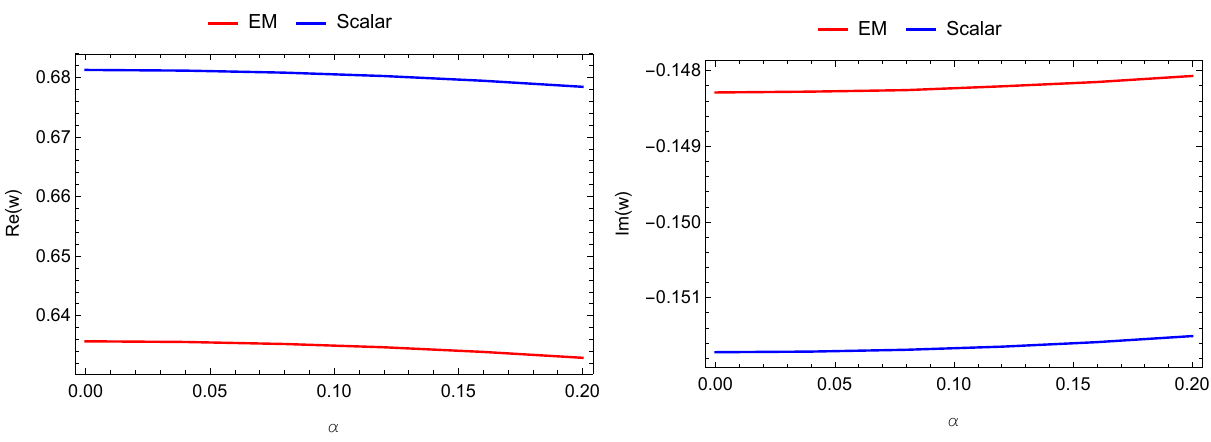}
    \caption{The real (left) and imaginary (right) parts of the quasinormal frequencies for both perturbations versus $\alpha$. Here, $l=2$}
    \label{figA1}
\end{figure}

\begin{figure}[ht!]
    \centering
    \includegraphics[scale=0.7]{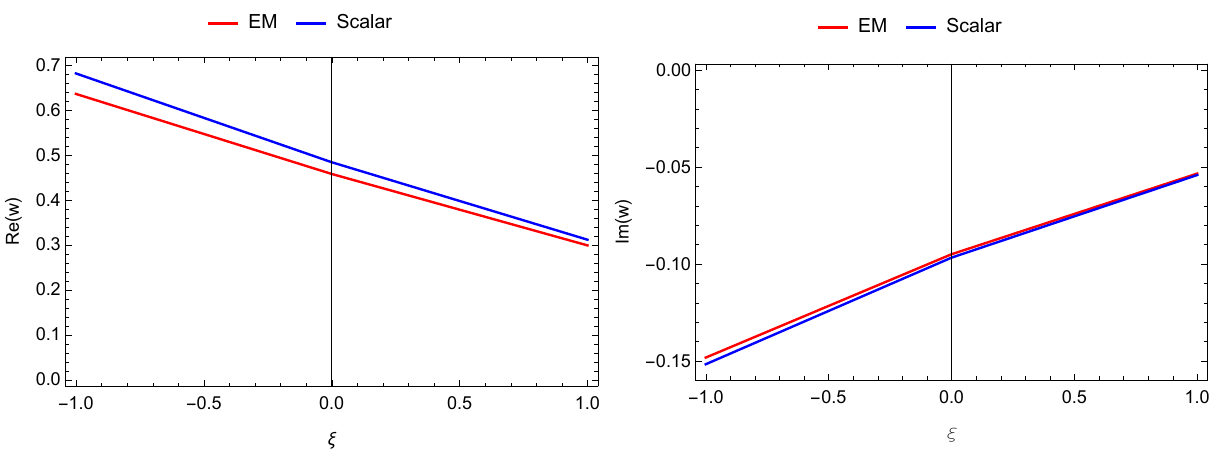}
    \caption{The real (left) and imaginary (right) parts of the quasinormal frequencies for both perturbations versus $\xi$. Here, $l=2$}
    \label{figZ1}
\end{figure}

As shown in Tables  \ref{taba1} and \ref{taba2}, the imaginary part is negative in all quasinormal frequency modes, suggesting that the quantum-corrected BH with Phantom Global Monopoles return to steady state following perturbation. In Figures \ref{figA1} and \ref{figZ1}, the real and imaginary quasinormal frequencies are plotted against the quantum correction parameter and the Monopole parameter to show how these parameters affect the quasinormal 
 spectrum. When the monopole parameter $\xi$ is constant, figures show that the real part of quasinormal frequencies, which corresponds to the frequency of gravitational waves, decreases as $\alpha$ increases for both forms of perturbations. On the other hand, the imaginary part of quasinormal frequencies has the opposite effect. This indicates that both the frequency of gravitational waves and the damping rate grow with $\alpha$. When $\alpha$ remains constant, both types of perturbations show comparable tendencies of increasing $\xi$. However, $\xi$ has a greater impact on both perturbations than parameter $\alpha$.  The system generally oscillates weakly and decays slowly.
 
\section{Conclusions} \label{sec5}

In this study, we analyzed the geodesic motion and the RW potential for a quantum-corrected AdS-Schwarzschild BH modified by the presence of phantom and ordinary global monopoles. The spacetime was described by a spherically symmetric metric incorporating quantum corrections and monopole effects, which substantially altered the BH's structure and dynamics.

We began by deriving the effective potential governing the motion of test particles. The analysis revealed that the potential was significantly influenced by the nature of the monopoles and the quantum correction parameter $\alpha$. For ordinary monopoles ($\xi = 1$), both the horizon radius and photon sphere radius were found to be larger compared to their counterparts in Schwarzschild and deformed-Schwarzschild metrics. For phantom monopoles ($\xi = -1$), these radii were smaller. These relationships were expressed as $r_{h} > r_{h}^{\text{deformed-Sch}} > r_{h}^{\text{Sch}}$ for ordinary monopoles and $r_{h} < r_{h}^{\text{Sch}} < r_{h}^{\text{deformed-Sch}}$ for phantom monopoles. Figures \ref{fig:13} and \ref{fig:14} illustrated the effective potential for null and time-like geodesics, showing the distinct behaviors induced by the two types of monopoles. These findings highlighted the role of quantum corrections in shaping the dynamics of test particles.

The study of the potential governing perturbations extended this analysis. Spin-dependent potentials were examined for scalar, electromagnetic, and gravitational fields. The results demonstrated that the potential depends on $\xi$, $\alpha$, the cosmological constant $\Lambda$, and the multipole number $\ell$. The potential for phantom monopoles was consistently higher than for ordinary monopoles, as evident from $\mathrm{b} > \mathrm{a}$. Figures \ref{fig:1}–\ref{fig:9} depicted the behavior of the potential for various spins and highlighted the substantial effect of quantum corrections. QNMs were investigated using the sixth-order WKB approximation for scalar and electromagnetic perturbations. An increase in $\alpha$ reduced the oscillation frequency and increased the damping rate, indicating enhanced stability. The monopole parameter $\xi$ had a more pronounced effect compared to $\alpha$, as shown in Figures \ref{figA1} and \ref{figZ1}. Tables \ref{taba1} and \ref{taba2} summarized the frequencies, providing insight into how quantum corrections and monopole effects influence the system's return to equilibrium after perturbations. 
Overall, this study highlighted the profound influence of quantum corrections and the topological effects of global monopoles on the physical characteristics of BHs. By meticulously analyzing these combined factors, we demonstrated how quantum modifications reshape the spacetime geometry and dynamics, while the presence of global monopoles introduces additional complexities that significantly impact the behavior of test particles and field perturbations. We believe that this work serves as one of the stepping stones for further explorations into the interplay between fundamental quantum effects and topological defects in astrophysical and cosmological contexts.

Future research could explore the thermodynamic implications of the quantum corrections and monopoles, such as corrections to entropy and temperature. Investigating rotating BHs or higher-dimensional spacetimes with these modifications could uncover further phenomena. Numerical studies of QNMs for lower multipole moments or beyond the WKB approximation \cite{Konoplya:2019hlu} may yield more precise results. Additionally, examining potential observational signatures, including gravitational wave patterns and lensing phenomena \cite{Kim:2020xkm}, could bridge theoretical predictions with astrophysical observations.

\section*{Conflict of Interests} 

Author(s) declares no such conflict of interests.

\section*{Data Availability Statement} 

No data were generated or analyzed in this study.

\section*{Acknowledgements}

F.A. acknowledges the Inter University Centre for Astronomy and Astrophysics (IUCAA), Pune, India for granting visiting associateship. \.{I}.~S. thanks T\"{U}B\.{I}TAK, ANKOS, and SCOAP3 for funding and acknowledges the networking support from COST Actions CA22113 and CA21106.


\begin{thebibliography}{}

\bibitem{ligo} B. P. Abbott {\it et al.}, Phys. Rev. Lett. {\bf 116}, 131103 (2016).

\bibitem{AA3} K. Akiyama {\it et al.}, Astrophys. J. {\bf 875}, L1 (2019) ; Astrophys. J. {\bf 875}, L2 (2019); Astrophys. J. {\bf 875}, L3 (2019); Astrophys. J. {\bf 875}, L4 (2019); Astrophys. J. {\bf 875}, L5 (2019); Astrophys. J. {\bf 875}, L6 (2019);  Astrophys. J. Lett. {\bf 930}, L12 (2022).

\bibitem{Hawking:1974rv} S.~W.~Hawking, Nature \textbf{248}, 30 (1974).

\bibitem{Gibbons:1977mu} G.~W.~Gibbons and S.~W.~Hawking, Phys. Rev. D \textbf{15}, 2738 (1977).

\bibitem{Hawking:1982dh} S.~W.~Hawking and D.~N.~Page, Commun. Math. Phys. \textbf{87}, 577 (1983).

\bibitem{Parikh:1999mf} M.~K.~Parikh and F.~Wilczek, Phys. Rev. Lett. \textbf{85}, 5042 (2000).

\bibitem{Sakalli:2017ewb} I.~Sakalli and A.~Ovgun, EPL \textbf{118}, 60006 (2017).

\bibitem{Sakalli:2015jaa} I.~Sakalli and A.~\"Ovg\"un, Gen. Rel. Grav. \textbf{48}, 1 (2016).
 
\bibitem{Ryu:2006bv} S.~Ryu and T.~Takayanagi, Phys. Rev. Lett. \textbf{96}, 181602 (2006).

\bibitem{Strominger:1996sh} A.~Strominger and C.~Vafa, Phys. Lett. B \textbf{379}, 99 (1996).

\bibitem{Bardeen:1973gs} J.~M.~Bardeen, B.~Carter and S.~W.~Hawking, Commun. Math. Phys. \textbf{31}, 161 (1973).

\bibitem{Eisert:2008ur} J.~Eisert, M.~Cramer and M.~B.~Plenio, Rev. Mod. Phys. \textbf{82}, 277 (2010).

\bibitem{Bousso:2002ju} R.~Bousso, Rev. Mod. Phys. \textbf{74}, 825 (2002).

\bibitem{ph1} R. Penrose, Phys. Rev. Lett. {\bf 14}, 57 (1965).

\bibitem{ph2} S. Hawking, R. Penrose, Proc. Roy. Soc. Lond. {\bf A. 314}, 529 (1970).

\bibitem{ph3}I. Dymnikova, Gen, Relat, Gravit. {\bf 24}, 235 (1992).

\bibitem{DIK} D. I. Kazakov, S. N. Solodukhin, Nucl. Phys. {\bf B 429}, 153, (1994).

\bibitem{ph5} I. Dymnikova, Class. Quantum Grav. {\bf 19}, 725 (2002).

\bibitem{ph6}S. G. Ghosh, Eur. Phys. J. C {\bf 75}, 532 (2015).

\bibitem{ph7} H. Culetu, Int. J. Theor. Phys. {\bf 54}, 2855 (2015).

\bibitem{ph8} A. Simpson, M. Visser, JCAP {\bf 02}, 042 (2019).

\bibitem{ph9} S. Upadhyay, Nadeem-ul-islam, P. A. Ganai, JHAP {\bf 2}, 25 (2022).

\bibitem{ph10} S. Gangopadhyay, S. Sen, R. Mandal, EPL {\bf 141}, 49001 (2023).

\bibitem{ph12} V. P. Frolov et al., Spherically symmetric collapse in quantum gravity, Phys. Lett {\bf 106}, 307 (1981).

\bibitem{ph13} T. Biswas, E. Gerwick, T. Koivisto, and A. Mazumdar, Phys. Rev. Lett. {\bf 108}, 031101 (2012).

\bibitem{ph11} M. J. Duff, Phys. Rev. {\bf D 9}, 1837 (1974).

\bibitem{CC1} R. A. Konoplya and A. Zhidenko. Rev. Mod. Phys. {\bf 83}, 793 (2011).

\bibitem{CC2} K. D. Kokkotas and B. G. Schmidt, Living Rev. Relativ. {\bf 2}, 2 (1999) 

\bibitem{Chen:2021cts} C.~Y.~Chen, M.~Bouhmadi-L\'opez and P.~Chen, Eur. Phys. J. Plus \textbf{136}, 253 (2021).

\bibitem{MB2} M. Bouhmadi-L’pez, S. Brahma, C.-Y. Chen, P. Chen and D. Yeom, JCAP 07 ({\bf 2000}) 066.

\bibitem{Moderski:2001tk} R.~Moderski and M.~Rogatko, Phys. Rev. D \textbf{64}, 044024 (2001)

\bibitem{CC3} S. Chandrasekhar and S. Detweiler, Proc. R. Soc. {\bf A 344}, 441 (1975) 

\bibitem{CC4} S. R. Dolan and A. C. Ottewill, Class. Quantum Grav. {\bf 26}, 225003 (2009). 

\bibitem{CC5} S. R. Dolan, Phys. Rev. {\bf D 82}, 104003 (2010). 

\bibitem{CC6} k. Lin, j. Li and n. Yang, Gen. Relativ. Gravit. {\bf 43}, 1889 (2011).

\bibitem{CC7} S. Mahapatra, J. High Energy Phys. {\bf 04}, 142 (2016).

\bibitem{CC8} H. J. Blome and B. Mashhoon, Phys. Lett. A {\bf 110}, 231 (1984).

\bibitem{CC9} E. Leaver, Proc. R. Soc. {\bf A 402}, 285 (1985). 

\bibitem{CC10} H. P. Nollert, Phys. Rev. {\bf D 47}, 5253 (1993).

\bibitem{CC11} G. T. Horowitz and V. E. Hubeny, Phys. Rev. {\bf D 62}, 024027 (2000). 

\bibitem{CC12} H. T. Cho {\it et. Al.}, Adv. Math. Phys. {\bf 2012}, 281705 (2012). 

\bibitem{CC13}  B. F. Schutz and C. M. Will , Astrophys. J. {\bf 291}, L33 (1985).

\bibitem{CC14} S. Iyer and C. M. Will, Phys. Rev. {\bf D 35}, 3621 (1987).

\bibitem{CC15} R. A. Konoplya,  Phys. Rev. {\bf D 68}, 024018 (2003).

\bibitem{Zahid:2024hwi} M.~Zahid, O.~Yunusov, C.~Shen, J.~Rayimbaev and S.~Muminov, Phys. Dark Univ. \textbf{47}, 101734 (2025).

\bibitem{Balart:2024rtj} L.~Balart, G.~Panotopoulos and \'A.~Rinc\'on, Ann. Phys. (NY) \textbf{473}, 169865 (2025).

\bibitem{Al-Badawi:2024jnt} A.~Al-Badawi, Y.~Sekhmani, J.~Rayimbaev and R.~Myrzakulov, Int. J. Mod. Phys. D \textbf{33}, 2450043 (2024)

\bibitem{Al-Badawi:2024cby} A.~Al-Badawi, S.~Shaymatov, S.~K.~Jha and A.~Rahaman, Eur. Phys. J. C \textbf{84}, 722 (2024).

\bibitem{Al-Badawi:2024iqv} A.~Al-Badawi and S.~K.~Jha, Commun. Theor. Phys. \textbf{76}, 095403 (2024).

\bibitem{Al-Badawi:2024iax} A.~Al-Badawi, S.~K.~Jha and A.~Rahaman, Eur. Phys. J. C \textbf{84}, 145 (2024).

\bibitem{Sasaki:2016jop} M.~Sasaki, T.~Suyama, T.~Tanaka and S.~Yokoyama, Phys. Rev. Lett. \textbf{117}, 061101 (2016). 

\bibitem{SCJJ} S. Chen, J. Jing, Class. Quantum Grav. {\bf 30}, 175012 (2013).

\bibitem{AHEP} M. Sharif, S. Iftikhar, Adv. High Energy Phys. {\bf 2015}, 854264 (2015).

\bibitem{MB} M. Barriola, A. Vilenkin, Phys. Rev. Lett. {\bf 63}, 341 (1989).

\bibitem{ERBM} E. R. Bezerra de Mello, Braz.J.Phys. {\bf 31}, 211 (2001).

\bibitem{wu} S. Wu, C. Liu, Nucl. Phys. B {\bf 985}, 115987 (2022).

\bibitem{Birrell:1982ix} N.~D.~Birrell and P.~C.~W.~Davies, \textit{Quantum Fields in Curved Space}, (Cambridge University Press, Cambridge, 1982).

\bibitem{Sakalli:2022xrb} \.I.~Sakalli and S.~Kanzi, Turk. J. Phys. \textbf{46}, 51 (2022).

\bibitem{AFA} A. F. Ali and M. M. Khalil, Nucl. Phys. {\bf B 909}, 173 (2016). 

\bibitem{TR} T. Regge, and J. A. Wheeler, Phys. Rev. {\bf 108}, 1063 (1957).

\bibitem{FJZ} F. J. Zerilli, Phys. Rev. Lett. {\bf 24}, 737 (1970).

\bibitem{Iyer} S. Iyer, C.M. Will, Phys. Rev. {\bf D 35}, 3621 (1987).

\bibitem{Konoplya} R. A. Konoplya, Phys. Rev. {\bf D 68}, 024018 (2003).

\bibitem{Konoplya:2019hlu} R.~A.~Konoplya, A.~Zhidenko and A.~F.~Zinhailo, Class. Quant. Grav. \textbf{36}, 155002 (2019).

\bibitem{Kim:2020xkm} K.~Kim, J.~Lee, R.~S.~H.~Yuen, O.~A.~Hannuksela and T.~G.~F.~Li, Astrophys. J. \textbf{915}, 119 (2021).

\end{thebibliography}
\end{document}